\documentclass[twocolumn]{aastex7}
\usepackage{amsmath}
\usepackage{mathtools}
\usepackage{multirow}

\graphicspath{{./}{figures/}}

\newcommand{\lya}{\ensuremath{\textrm{Ly}\alpha}}
\newcommand{\ewlya}{\ensuremath{\textrm{EW}_{\textrm{Ly}\alpha}}}

\newcommand{\sub}[1]{\ensuremath{_\textrm{\scriptsize{#1}}}}
\newcommand{\arcs}{\ensuremath{''}}

\newcommand{\msun}{\ensuremath{\rm M_\odot}}

\defcitealias{trainor2013}{T13}
\defcitealias{trainor2015}{T15}
\defcitealias{trainor2016}{T16}

\shorttitle{Lya Halos at $z\sim 2-3$}
\shortauthors{Trainor et al.}

\begin{document}

\title{The Lyman-alpha Halos of Galaxies at $z \approx 2-3$ in the Keck Baryonic Structure Survey}
\author[0000-0002-6967-7322]{Ryan F. Trainor}
\affiliation{Department of Physics \& Astronomy, Franklin \& Marshall College, 415 Harrisburg Pike, Lancaster, PA 17603}
\affiliation{William H. Miller III Department of Physics and Astronomy, Johns Hopkins University, Baltimore, MD 21218, USA}
\email{ryan.trainor@fandm.edu}

\author[0000-0002-2052-822X]{Noah R. Lamb}
\affiliation{Department of Physics \& Astronomy, Franklin \& Marshall College, 415 Harrisburg Pike, Lancaster, PA 17603}
\affiliation{Department of Physics, Drexel University, 32 S. 32nd Street
Philadelphia, PA 19104}
\email{noah.r.lamb@drexel.edu}

\author[0000-0002-2052-822X]{Charles C. Steidel}
\affiliation{Cahill Center for Astrophysics, California Institute of Technology, MC 249-17, 1200 East California Boulevard, Pasadena, CA 91125, USA}
\email{ccs@astro.caltech.edu}

\author[0000-0003-4520-5395]{Yuguang Chen}
\affiliation{Cahill Center for Astrophysics, California Institute of Technology, MC 249-17, 1200 East California Boulevard, Pasadena, CA 91125, USA}
\affiliation{Department of Physics, The Chinese University of Hong Kong, Shatin, N.T., Hong Kong SAR, China}
\email{yuguangchen@cuhk.edu.hk}

\author[0000-0001-9714-2758]{Dawn K. Erb}
\affiliation{The Leonard E. Parker Center for Gravitation, Cosmology and Astrophysics, Department of Physics,
University of Wisconsin-Milwaukee, 3135 N Maryland Avenue, Milwaukee, WI 53211, USA}
\email{erbd@uwm.edu}

\author[0000-0003-4941-2201]{Elizabeth Trenholm}
\affiliation{School of Computing and Mathematical Sciences, University of Greenwich, London SE10 9LS, UK}
\affiliation{Department of Astronomy, University of California, Berkeley, 501 Campbell Hall, Berkeley, CA 94720, USA}
\email{etrenholm@berkeley.edu}

\author[0000-0002-6187-4866]{Rebecca L. McClain}
\affiliation{Department of Physics \& Astronomy, Franklin \& Marshall College, 415 Harrisburg Pike, Lancaster, PA 17603}
\affiliation{Department of Astronomy, The Ohio State University, 140 West 18th Avenue, Columbus, OH 43210, USA}
\affiliation{Center for Cosmology and Astroparticle Physics (CCAPP), 191 West Woodruff Avenue, Columbus, OH 43210, USA}
\email{mcclain.378@buckeyemail.osu.edu}

\author[0009-0005-2278-0794]{Io Kovach}
\affiliation{Department of Physics \& Astronomy, Franklin \& Marshall College, 415 Harrisburg Pike, Lancaster, PA 17603}
\affiliation{Illinois Center for Advanced Studies of the Universe \& Department of Physics, University of Illinois at Urbana-Champaign, Urbana, Illinois 61801, U.S.A.}
\email{iok2@illinois.edu}

% \correspondingauthor{Noah R. Lamb}
% \email{noah.r.lamb@drexel.edu}
\correspondingauthor{Ryan Trainor}
\email{ryan.trainor@fandm.edu}

\begin{abstract}
 We present the large-scale spatial \lya\ profiles of galaxies from the Keck Baryonic Structure Survey (KBSS) at $2<z<3$. This work also describes the \lya\ imaging for the KBSS-\lya\ survey for the first time. Our sample includes 734 \lya-selected galaxies and 119 continuum-selected galaxies with \lya\ narrow-band imaging, and we measure the spatial morphology of \lya\ and continuum emission for stacked subsamples of these two populations. These samples allow us to study the variation in \lya\ emission profiles over a broad range of UV continuum luminosities and \lya\ equivalent widths (EW$_{\mathrm{Ly}\alpha}$), including systems with net \lya\ absorption in slit spectroscopy. We characterize the spatial profiles using two techniques: directly fitting an exponential function to the stacked profile, and a multi-component forward-modeling technique using the empirical large-scale PSF. We find that both methods yield similar results and that the forward-modeling technique self-consistently fits profiles exhibiting central \lya\ emission or \lya\ absorption, with the spatial scale of central \lya\ approximately matching that of the continuum emission. We also find extended \lya\ emission such that all our subsamples---including central \lya\ absorbers---are net \lya\ emitters on scales comparable to the circumgalactic medium ($R\gtrsim 50\,\textrm{kpc}$, $\theta\gtrsim 6''$). We find that the scale length of the \lya\ halo is not strongly dependent on the properties of the central galaxy, including its net continuum luminosity or EW$_{\mathrm{Ly}\alpha}$, although we find a possible weak tendency of continuum-faint, high-EW$_{\mathrm{Ly}\alpha}$ galaxies to exhibit larger \lya\ halos in contrast with previous work.
\end{abstract}

\section{Introduction}
\label{sec:intro}

A key frontier for understanding galaxy formation is the circumgalactic medium (CGM), the gaseous region around galaxies that is shaped by an interplay of accretion, expulsion, and ionization. The CGM is distinct from the interstellar medium (ISM) in that it describes the gas at larger galactocentric radii than the bulk stellar population ($r\gg 1\,$ kpc). 
The outer radius of the CGM is often defined by the $r_\textrm{vir}\sim 100$ kpc virial radius of a galaxy's dark matter halo (e.g., \citealt{tumlinson2017}), although the local properties of the intergalactic medium (IGM) are highly correlated with the presence of a nearby galaxy to even larger distances ($r\gtrsim1$ Mpc; e.g., 
\citealt{adelberger2003,adelberger2005,rudie2012a,chen2020}).

Observationally, the CGM is a multiphase medium with complex and varied kinematics, temperature, ionization, and metallicity, as confirmed by absorption-line measurements in the local Universe (e.g., \citealt{tumlinson2011,tumlinson2017,werk2014,werk2016}),  at intermediate redshifts ($z\sim 0.5-1$; e.g., \citealt{chen2018,cooper2021}), and at early cosmic times ($z\gtrsim2$; \citealt{rudie2012a,rudie2019,steidel2010,lau2016,lau2018,chen2020}). Using theoretical models that reproduce this complexity (e.g., \citealt{shen2013,ford2013,suresh2015,fielding2017,hummels2019,schneider2020}), the CGM can in principle be used to study the exchange of gas between galaxies and the intergalactic medium, feedback effects from the stars and active galactic nuclei (AGN) within galaxies, and the external ionization of the UV background \citep{haardt2012} or nearby luminous AGN \citep{cantalupo2012,trainor2013,cantalupo2017}. While absorption-line spectroscopy provides exquisite constraints on gas kinematics and ionization along individual lines of sight through the CGM, studying the CGM in emission has the potential to provide a spatially-resolved view of the processes that sculpt galaxies throughout their evolution.

At high redshifts, extended hydrogen \lya\ emission ($\lambda_\textrm{rest}=1216$\AA) is a particularly powerful tool for studying the CGM in emission. Typically the most luminous emission line in the rest-UV spectra of star-forming galaxies, \lya\ provides a bright tracer of photoionized gas that does not rely on metal enrichment and falls within the observed optical wavelength range at the peak redshifts of star formation, AGN activity, and their associated feedback ($z\approx 2-3$; \citealt{madau2014}). In addition, diffuse \lya\ emission has been found to surround star-forming galaxies, with these \lya\ ``halos'' being detected in stacked narrow-band images of typical high-redshift galaxies (\citealt{steidel2011,matsuda2012,momose2014,momose2016,kikuta2023}) or stacked integral-field (IFU) observations \citep{lujanniemeyer2022}. More recently, similar \lya\ halos have now been measured in individual galaxies at low redshifts \citep{hayes2013,hayes2014,rasekh2021,runnholm2023} and at $z>2$ \citep{wisotzki2016,leclercq2017,leclercq2020,chen2021,erb2023} using deep IFU spectroscopy.

Notably, not all galaxies exhibit strong \lya\ emission: variations in the rate of photon production as well as the complex radiative transfer processes that regulate their later escape may cause individual galaxies to appear as net \lya\ emitters or absorbers, particularly when observed on small spatial scales as in down-the-barrel spectra.\footnote{See \citet{trainor2019} and \citet{runnholm2020} for measurements and discussion of these effects at high and low redshifts, respectively, as well as a recent review of \lya\ observations by \citet{ouchi2020}} However, even galaxies without central \lya\ emission are often seen to produce net \lya\ emission on larger scales \citep{steidel2011,kusakabe2022}, and the near-ubiquity of \lya\ halos in searches of sufficient depth (SB(\lya)~$\approx$~10$^{-19}$ erg s$^{-1}$ cm$^{-2}$ arcsec$^{-2}$) indicates that they arise from processes that occur generically within star-forming galaxies and/or their surrounding CGM.

Many different physical processes may be capable of producing these \lya\ halos: resonant scattering of \lya\ photons that originate in star-forming regions \citep{laursen2007,steidel2011,zheng2011}, star-formation in satellite galaxies \citep{momose2016,mas-ribas2017a,kikuta2023}, gravitational cooling radiation \citep{haiman2000,rosdahl2012}, and ``fluorescent'' recombination from photoionization within the CGM \citep{gallego2018,gallego2021,mas-ribas2016} have all been proposed, as well as other models or combinations of the above (e.g., \citealt{lake2015,mitchell2021}). Recent studies have attempted to constrain these models by comparing properties of the \lya\ halos with those of their host galaxies. However,  even extremely sensitive observations of individual \lya\ halos with VLT/MUSE \citep{bacon2010} and Keck/KCWI \citep{morrissey2018} around galaxies with well-characterized stellar populations have failed to demonstrate any conclusive relationships between these galaxies and their halos \citep{wisotzki2016,leclercq2017,chen2021}.

Despite the sensitivity to individual halos that has been afforded by IFU spectroscopy, a significant drawback of recent galaxy surveys for studying the variation in \lya\ halo properties is the fact that many of these surveys sample a relatively narrow range of the parameter space of star-forming galaxies. In particular, the samples of \citet{wisotzki2016} and \citet{leclercq2017} consist of \lya-emitters (LAEs) selected on the basis of their prominent central \lya\ emission, such that these galaxies are unlikely to reflect the full range of processes by which \lya\ escapes from galaxies. Conversely, \citet{chen2021} included both \lya-selected and continuum-selected galaxies from the Keck Baryonic Structure Survey (KBSS; \citealt{rudie2012a,steidel2014,strom2017}). That work measured the association between stellar and halo morphologies but did not attempt to characterize the variation of \lya\ halo properties with observables such as galaxy luminosity and central \lya\ emission.

Earlier work by \citet{steidel2011} analyzed 92 galaxies of which approximately half exhibit central \lya\ absorption. As noted within that work, \citet{steidel2011} did not attempt to correct for the terrestrial PSF in their estimations of galaxy and halo parameters, whereas several more recent estimates such as the VLT/MUSE surveys described above have included these factors and found significantly smaller halos. The different techniques and results of these studies are difficult to compare directly, and \citet{feldmeier2013} speculate that accounting for the large-scale PSF would significantly alter the results of \citet{steidel2011}.

Most recently, \citet{kusakabe2022} used VLT/MUSE to study the incidence of \lya\ halos around a sample of UV-continuum-selected galaxies at $z\approx 3-6$, finding a similar halo fraction and radial profile as in \lya-selected samples. However, that sample is relatively small ($N=21$), with the majority of objects falling in a narrow range of UV luminosities ($-18.5\lesssim M_{1500}\lesssim-19.5$), so it again has significant limitations for studying the broad parameter space of galaxy properties. In addition, the inability of VLT/MUSE to measure \lya\ at $z\lesssim3$ also requires complementary galaxy samples to understand any variation in halo properties with redshift.

This paper seeks to characterize the \lya\ halo profiles of star-forming galaxies at $z\approx2-3$ across a broad range of UV continuum luminosities and central \lya\ emission.  Our analysis includes 853 galaxies drawn from the KBSS and KBSS-\lya\ 
% \citep{trainor2015,trainor2016}
(\citealt{trainor2015,trainor2016}, hereafter \citetalias{trainor2015} and \citetalias{trainor2016})
using both UV continuum selection (119 galaxies) and \lya\ selection (734 galaxies) in order to maximize the diversity of galaxy properties included in our analysis. Furthermore, we fit the \lya\ halo profiles using multiple techniques in order to facilitate comparison with previous studies including \citet{steidel2011} and those conducted with VLT/MUSE, thereby enabling consistent comparison of halo profiles across a large space of galaxy properties, environments, and cosmic epochs.

The paper is organized as follows. Sec.~\ref{sec:obs} describes our broadband and narrowband imaging observations, which are used to construct the KBSS-\lya\ sample of \lya-emitters and to measure the \lya\ profiles of both the KBSS and KBSS-\lya\ samples. Sec.~\ref{sec:analysis} describes our analysis of the raw images, including the creation of images that isolate the stellar continuum emission and \lya\ line emission, and the construction of stacked images for subsamples of the full KBSS and KBSS-\lya\ samples. Sec.~\ref{sec:halos} describes measurements of the \lya\ halo profiles, which we fit in two ways: (1) via direct exponential fits to the extended \lya\ profiles as in \citet{steidel2011}, and (2) using a forward-modeling approach similar to that of \citet{leclercq2017} and other recent work that treats the central ``galaxy'' and large scale ``halo'' as individual exponential terms that are then convolved with the empirical point-spread function (PSF). In Sec.~\ref{sec:comparison}, we compare the results of our two model frameworks and discuss relationships among the measured galaxy and halo properties, including their sizes, luminosities, and central \lya\ emission. Sec.~\ref{sec:discussion} compares our results to other recent work in the literature, and Sec.~\ref{sec:summary} summarizes our conclusions. Details on the construction of the large-scale PSF used for the forward modeling as well as additional results from the forward-modeling fits are presented in the appendices. Throughout the paper, we assume a Planck 2018 cosmological model \citep{aghanim2020} with $H_0=67.4\,\textrm{km}\,\textrm{s}^{-1}\,\textrm{Mpc}^{-1}$ and $\Omega_m=1-\Omega_\Lambda=0.315$. Distances are given in physical kiloparsecs (kpc) unless otherwise noted, and all magnitudes are given in the AB system \citep{oke1983}.

\section{Observations} %name?
\label{sec:obs}
\subsection{Imaging and Data Reduction}
\label{sec:imaging}
The data presented in this work were collected over the course of the
KBSS-\lya\ survey previously described by \citet[hereafter T13]{trainor2013},
\citetalias{trainor2015}, and \citetalias{trainor2016}
% and \citet{trainor2015,trainor2016}, 
but the imaging data and photometry
for that survey are presented here for the first time.\footnote{In
  \citetalias{trainor2015}, the KBSS-\lya\ imaging observations are
  described as being presented by ``R. Trainor et al. (2015, in
  preparation),'' which has now been subsumed into this paper. We apologize for the delay.} The KBSS-\lya\ comprises nine survey fields, each centered on a bright QSO at $z\approx2.5-2.8$ (Table~\ref{table:fields}). We conducted deep imaging in each survey field using custom
narrowband (NB) filters that cover the wavelength of \lya\ at the redshift of the corresponding QSO , and we also obtained broadband images using filters sampling the
continuum near \lya. These observations were obtained over a period of several
years (2007-2015) using Keck 1/LRIS \citep{oke95,steidel2004} and are described in
Table~\ref{table:fields}. The images described in this work were collected with the UV/blue-optimized LRIS-B spectrograph channel described by \citet{steidel2004} using the D500 or D560 dichroics. All nine fields are part of the Keck Baryonic Structure Survey (KBSS;
\citealt{rudie2012a}) and hence have ancillary images in a variety of
broadband filters; the collection and reduction of these data are
described in \citet{steidel2004}, \citet{erb06a}, \citet{reddy2008}, \citet{reddy2012}, and references 
therein. 

\begin{deluxetable*}{lccccccc}
\tablecaption{Fields Descriptions \& Observations}
\tablewidth{0pt}
\tablehead{
\colhead{QSO Field} & $z\sub{Q}$ ($\pm$0.001) & NB Filter & FWHM (\AA)
& Exp (s) & BB Filter & $N_\textrm{KBSS}$ & $N_\textrm{KBSS-Ly$\alpha$}$
}\colnumbers
\startdata
Q0100+13 (PHL957)   &  $ 2.721 $ &   NB4535  & 76 & 24,590 & B+G & 15 &  62 \\
HS0105+1619         &  $ 2.652 $ &   NB4430  & 72 & 21,600 & B   & 9  &  68\\
Q0142$-$10 (UM673a) &  $ 2.743 $ &   NB4535  & 76 & 18,000 & B+G & 5  &  39 \\
Q1009+29 (CSO 38)   &  $ 2.652 $ &   NB4430  & 72 & 18,000 & B   & 17 &  63 \\
HS1442+2931         &  $ 2.660 $ &   NB4430  & 72 & 18,000 & B   & 13 & 125\\
HS1549+1919         &  $ 2.843 $ &   NB4670  & 88 & 18,000 & G   & 29 & 178 \\
HS1603+3820 & $2.551$ & NB4325 & 74 & 10,800  & B   & 9  &  66\\
HS1700+6416         &  $ 2.751 $ &   NB4535  & 76 & 23,400 & B+G & 9  &  55\\
Q2343+12            &  $ 2.573 $ &   NB4325  & 74 & 15,896 & B   & 15 &  78 
\enddata
\tablecomments{Each NB filter name (3) corresponds approximately to the
  central wavelength of the filter in \AA, and the NB filter widths are given in column (4). (5) is the total exposure time on LRIS-B with the
  corresponding NB filter, and (6) refers to the broadband filter (${B}$, ${G}$,
 or a combination of the two) used to measure the continuum magnitude
of the LAEs in each field (Sec.~\ref{sec:imaging}). Filter details are available from the public Keck/LRIS website. Columns (7) \& (8) give the total number of KBSS (continuum-selected and spectroscopically-confirmed) and KBSS-\lya\ (\lya-selected) galaxies in each field that fall within the NB filter bandpass. Galaxy selection is described in Secs.~\ref{sec:selection}~\&~\ref{sec:kbss-sample}.}
\label{table:fields}
\end{deluxetable*}

The NB images are taken as a series of 1800 s exposures with a dither
pattern optimized to cover the  13.5\arcs chip gap on LRIS-B. Each
field had $\sim$10 NB exposures for a total of $\sim$5 hours of observing
time per field, for a typical 3$\sigma$ depth of
$m\sub{NB}(3\sigma)\sim 26.7$ as estimated through Monte-Carlo
simulations of fake objects. The recovered fraction of
injected objects in these simulations suggest that our survey is
$>$95\% complete (cumulatively) for $m\sub{NB}<26.8$ and has a
magnitude-specific completeness $\gtrsim$90\% for all magnitude bins
$m\sub{NB}\lesssim26.8$. We use 4 NB filters for this 
survey, and each field is imaged with the NB filter 
centered most closely on the wavelength of \lya\ at the redshift of its
corresponding QSO. Other than their central wavelengths, the 4 NB
filters have similar transmission properties, with FWHM $\sim 80$ \AA\ and a peak transmission $\sim$85\% in all cases. As the QSOs span a redshift range $2.573 \le z
\le 2.843$, the filter width corresponds to $\Delta z \approx 0.066$
or $\Delta v \approx 5400$ km s$^{-1}$ at their median
redshift. See Table~\ref{table:fields} for
more detailed statistics on each field and descriptions of the
individual NB filters.

We use LRIS-B $B$ and/or $G$ broadband filters to sample the continuum emission local to
\lya\ in each field, with the choice of filter depending on the corresponding NB filter in each field.
The continuum image
for each field is the ${B}$ image, ${G}$ image, or an equal-weighted
combination of the two images (see Table~\ref{table:fields}). In each
field, the effective central wavelength of the continuum image is
within $\sim$60 \AA\ of that of the correponding NB filter. Broadband
images are taken with exposure times of $\gtrsim$1 hour, for a
typical 3$\sigma$ depth of $m\sub{BB}(3\sigma)\sim 28$, again
estimated through simulations of fake objects. Henceforth, $m\sub{BB}$
denotes the measured magnitude from the broadband
filter(s) used to infer the UV continuum value (${B}$, ${G}$, or their average).

We use standard IRAF routines to reduce the data via bias
subtraction, flat-fielding, cosmic-ray rejection, sky-subtraction, and
image registration and combination. Before the final image
combination, individual exposures are astrometrically registered to
LRIS-R $\mathcal{R}_s$-band images previously obtained as part of the KBSS and
corrected for distortion and rotation; the use of these deep
broadband images for image registration ensures a high density of
sources across the entire field for our astrometric alignment. In this
manner, we optimize the 
relative astrometric registration between our NB and \lya-continuum
images as well as their correspondence to the ancillary KBSS data. As
part of this process, all the images are registered to the 0\farcs211
pix$^{-1}$ scale of the pre-2009 LRIS-R detector\footnote{The LRIS-R detector was upgraded to match the 0\farcs135
pix$^{-1}$ scale of LRIS-B in June 2009, and many of the ancillary $\mathcal{R}_s$-band images were collected prior to the upgrade.} before being combined. An iterative smoothing process
is performed to one or both of the NB and \lya-continuum images in
order to match the PSFs of the two images and across fields; our PSF-matching procedure
is described in detail in Appendix~\ref{sec:psf}. In practice, small astrometric distortions remain between the NB and broadband filters in certain regions of the resulting wide-field images; these distortions are corrected locally during the construction of the \lya\ and continuum stamps as discussed in Sec.~\ref{sec:stamps}.

Photometric calibration of the $B$, $G$, and $\mathcal{R}_s$ images is performed
using photometric standard stars from the catalogs of
\citet{mas88} and \citet{oke1990} as described in \citet{steidel2003}. The NB
images are calibrated iteratively with 
respect to the broadband images during the selection of \lya-emitters  as described below.

\begin{deluxetable*}{l@{\hspace{1cm}}crcrcr@{}l}
%\footnotesize
\tablecaption{Stacked Galaxy Samples}
\tablewidth{0pt}
\tablehead{
\colhead{Sample} &
\colhead{Criterion}
& $N_\textrm{gal}$ &
$\langle M_\textrm{UV}\rangle$ 
&\multicolumn{2}{c}{$\langle \mathrm{EW}_{\textrm{Ly}\alpha}^\textrm{ind}/\mathrm{\AA}\rangle$}
&\multicolumn{2}{c}{$\langle \mathrm{EW}_{\textrm{Ly}\alpha}^\textrm{50}/\mathrm{\AA}\rangle$}
\\
\multicolumn{1}{c}{(1)}&\multicolumn{1}{c}{(2)}&\multicolumn{1}{c}{(3)}&\multicolumn{1}{c}{(4)}&\multicolumn{2}{c}{(5)}&\multicolumn{2}{c}{(6)}
}
\startdata
\multicolumn{4}{l}{\it KBSS-\lya\ galaxy samples}\\
All   &  $-$  &   734  & $-18.1$  & \phantom{aaa}40.2 &\phantom{a}& \phantom{aa}107&$_{-  7}^{+ 8}$\\
Faint & $M_\textrm{UV}>-18.1$  & 367 & $-17.5$ & $50.1$ && 178&$_{- 21}^{+33}$\\
Bright & $M_\textrm{UV}<-18.1$  & 367 & $-18.9$ & $29.7$ && 87&$_{-  6}^{+ 7}$\\
LoEW & $\textrm{EW}_{\textrm{Ly}\alpha}<40.2$\AA  & 367 & $-18.5$ & $22.8$ && 
56&$_{-  4}^{+ 7}$\\
HiEW & $\textrm{EW}_{\textrm{Ly}\alpha}>40.2$\AA  & 367 & $-17.8$ & $64.5$ && 196&$_{- 17}^{+19}$\\[5pt]\hline
\multicolumn{4}{l}{\it KBSS galaxy samples}\\
All   &  $-$  &   119  & $-20.7$  & 1.2  && $23$&$^{+3}_{-2}$\\
Abs & $F_{\textrm{Ly}\alpha}<0$  & 55 & $-20.8$ &$-6.0$ && $11$&$^{+7}_{-4}$\\
Em & $F_{\textrm{Ly}\alpha}>0$  & 64 & $-20.4$ & 10.8 && $43$&$^{+6}_{-5}$ \\
LAE & $\textrm{EW}_{\textrm{Ly}\alpha}>20$\AA  & 20 & $-20.2$ & 37.5 && $71$&$^{+11}_{-11}$\\
\enddata
\tablecomments{Measurements of the \lya\ flux $F_{\textrm{Ly}\alpha}$ and \lya\ equivalent width $\textrm{EW}_{\textrm{Ly}\alpha}$ for individual galaxies (i.e., columns 2 \& 5) are calculated within a 1\farcs2 diameter aperture, similar to the typical width of a Keck/LRIS slit. Conversely, $\langle \mathrm{EW}_{\textrm{Ly}\alpha}^\textrm{50}/\mathrm{\AA}
\rangle$ in column (6) represents the total \lya\ equivalent width measured from the stack of \lya\ and
continuum images using a 50 kpc $\approx$ 6$''$ radius.
Sec.~\ref{sec:curveofgrowth} discusses the variation in
$\text{EW}_{\text{Ly}\alpha}$ with aperture radius.}
\label{table:stacks}
\end{deluxetable*}

\subsection{Photometric selection of the KBSS-\lya\ sample}
 \label{sec:selection}
 
Source Extractor \citep{bertin1996}
is used in two-image mode to select objects in the NB image and
measure their magnitudes in both the NB and broadband images. For each
field, the resulting catalog and color-magnitude diagram contains a
well-defined ridge of points with constant
$m\sub{NB}-m\sub{BB}$; the zero-point of the NB image is then
adjusted iteratively until this ridge-line corresponds to
$m\sub{NB}-m\sub{BB}=0$. In fields where this ridge line is not
constant in $m\sub{NB}-m\sub{BB}$ color, a color-dependent
correction is applied to the broadband fluxes. In all cases, this
color correction effectively {\it increases} the broadband flux, a choice
we make to ensure a lack of contamination by low-EW$_{\lya}$ sources.
However, this choice may cause us to underestimate the
value of EW$_{\lya}$ for the faintest objects in our samples.

Spectroscopy was conducted for 318 of the photometrically-selected \lya-emitters using Keck/LRIS as described by \citetalias{trainor2015}. The success rate of our initial follow-up spectroscopy drops significantly
above $m\sub{NB}=26.5$ as noted in that paper, 
so we restrict the photometric LAE sample for this paper to those with
$22<m\sub{NB}<26.5$. The initial object selection used in
\citetalias{trainor2013} is based on the color excess
$m\sub{BB}-m\sub{NB}>0.6$ (corresponding closely to a
rest-frame equivalent width in \lya\ $[\text{EW}_{\lya}]\gtrsim 20$\AA). However,
this color excess is not a perfect proxy for equivalent width: the
range of redshifts of our QSO fields mean that a given value in EW$_{\lya}$
corresponds to a slightly different color for each narrowband filter.

We therefore calculate EW$_{\lya}$ for each object via the following assumptions: (1) both filters transmit 100\% of the flux within the FWHM of the effective wavelength midpoint of the band and 0\% of the flux outside, and (2) the continuum has a constant\footnote{Although $f_\lambda$ is not constant across the broadband filter in practice, note that the close similarities in the effective central wavelengths of our NB and BB filters means that linear terms in $f_\lambda$ will not affect the inferred value of \ewlya.} specific flux ($f_\mathrm{\lambda, cont}$) over both the NB and BB bandpasses. With these two approximations, we can express the measured flux in the NB and BB filters in terms of the underlying Ly$\alpha$ and continuum emission:
\begin{align}
F_\mathrm{NB}&=F_\mathrm{Ly\alpha}+f_\mathrm{\lambda,cont}\Delta_\mathrm{NB}\\
F_\mathrm{BB}&=F_\mathrm{Ly\alpha}+f_\mathrm{\lambda,cont}\Delta_\mathrm{BB}
\end{align}
where $F_\mathrm{NB}$ and $F_\mathrm{BB}$ are the integrated fluxes in the NB and BB filters, respectively. $F_\mathrm{Ly\alpha}$ is the Ly$\alpha$ line flux, and $f_\mathrm{\lambda, cont}$ is the   specific flux in the UV continuum, and $\Delta_\mathrm{NB}$ and $\Delta_\mathrm{BB}$ are the effective FWHM of the NB and BB\footnote{For fields which use the combined $B+G$ images for the continuum estimation, $\Delta_\textrm{BB}$ is the average of the $B$ and $G$ FWHMs.} filters (see Table~\ref{table:fields} for filter descriptions). Solving for $F_\mathrm{Ly\alpha}$ and $f_\mathrm{\lambda,cont}$, we have the following:
\begin{equation}
F_\mathrm{Ly\alpha}=\frac{F_\mathrm{NB}\Delta_\mathrm{BB}-F_\mathrm{BB}\Delta_\mathrm{NB}}{\Delta_\mathrm{BB}-\Delta_\mathrm{NB}}\label{eq:flya}%\\
\end{equation}
\begin{equation}
f_\mathrm{\lambda,cont}=\frac{F_\mathrm{BB}-F_\mathrm{NB}}{\Delta_\mathrm{BB}-\Delta_\mathrm{NB}}\label{eq:fcont}
\end{equation}
The rest-frame \lya\ equivalent width can then be calculated from the inferred \lya\ and continuum fluxes:
\begin{align}
\mathrm{EW}_{\mathrm{Ly}\alpha}&=\frac{1}{1+z_Q}\frac{F_\mathrm{Ly\alpha}}{f_\mathrm{\lambda,cont}}\label{eq:ewlya}
\end{align}

\noindent where $z\sub{}$ is the redshift of the nearby QSO in each field (Table~\ref{table:fields}), taken to be the
most likely redshift for the selected LAEs.\footnote{Note that \citetalias{trainor2015} demonstrate using rest-UV spectra that the majority of the LAEs are tightly clustered near the QSO redshift in each field, rather than filling the NB-selected volume uniformly.} \ewlya\ values are calculated for individual galaxies using a 1\farcs2 diameter aperture, similar to the width of a typical LRIS spectroscopic slit.

Many of the KBSS-\lya\ candidates are formally undetected in the BB images or have a BB flux consistent with the inferred \lya\ flux that falls within the bandpass of both the NB and BB filters. For these galaxies, only an upper limit on $f_\mathrm{\lambda,cont}$ and lower limit on EW$_{\lya}$ can be determined. In general, the uncertainty on EW$_{\lya}$ is highly asymmetric for strong \lya-emitters, so we calculate uncertainties and lower limits on EW$_{\lya}$ by perturbing the inferred values of $F_\textrm{NB}$ and $F_\textrm{BB}$ by a gaussian noise distribution with $\sigma$ given by the measurement uncertainty appropriate for each image. This procedure is repeated 1000 times for each object, with the 1$\sigma$ (2$\sigma$) lower limit on EW$_{\lya}$ taken to be the 32\%-tile (5\%-tile) value of EW$_{\lya}$. Selected galaxies are those for which 68\% of EW$_{\lya}$ values calculated in this manner are greater than 20\AA; in addition, the galaxy must be well-detected in the NB image ($m_\text{NB}<26.5)$ as described above.

As in \citetalias{trainor2013}, individual extended sources
(FWHM $>3\arcsec$) are removed from our sample for this
analysis; extended \lya\ sources (i.e., \lya\ ``blobs'' as in \citet{steidel2000}) in the KBSS fields are discussed elsewhere \citep{erb2011,martin2014,martin2015,martin2019,kikuta2019,osullivan2020}. We likewise exclude objects visually identified as likely contaminants
such as those near bright stars, image defects, or the edge of the
detector, as well as contaminants identified via spectroscopy. As described by \citetalias{trainor2015}, the number of low-redshift interlopers identified in the spectroscopic sample implies a low-redshift contamination rate of $<$3\% in our photometric sample; the majority of these contaminants are lower-redshift AGN with strong \ion{C}{4} or \ion{He}{2} emission.

In total, these 
criteria define our sample of 734 \lya-selected galaxies. As described by \citetalias{trainor2015}, these galaxies have typical rest-UV continuum magnitudes $\langle M_\textrm{UV}\rangle= -18.3$, which corresponds to $L\approx 0.1L_*$ at $z\sim2-3$ according to the galaxy luminosity functions of \citet{reddy2008}, and they have typical dynamical masses $\langle M_\textrm{dyn}\rangle=8\times10^8$~\msun\  determined from HST/WFC3-IR morphologies and MOSFIRE nebular velocity dispersions. $M_\textrm{UV}$ measurements are based on Source Extractor magnitudes from the LRIS-R $\mathcal{R}_s$-band images described above; these images are typically deeper than the $B$ and $G$ images used for our continuum subtraction and have the additional benefit of avoiding contamination by \lya. The $\mathcal{R}_s$ band has a central wavelength $\lambda_\text{obs}\approx6800$\AA, corresponding to $\lambda_\text{rest}\approx2000$\AA. The number of KBSS-\lya\ galaxies selected in each field is given in Table~\ref{table:fields} under the ``$N_\text{KBSS-Ly$\alpha$}$'' column.

\subsection{Selection of the KBSS sample}
\label{sec:kbss-sample}

In addition to the \lya-selected galaxy sample, we consider higher-luminosity galaxies from the KBSS that are selected by their rest-UV continuum colors (i.e., ``LBGs'' as described by \citetalias{trainor2015}). In particular, we choose the subset of KBSS galaxies with spectroscopic redshifts placing their \lya\ transitions within the NB passband for their corresponding field. This selection yields a sample of 121 $L\approx L_*$ galaxies ($M_\textrm{UV}\approx-21$). Notably, this latter sample of galaxies is similar to and overlaps with the sample of galaxies described by \citet{steidel2011}, which allows us to revisit the \lya-halo measurements of that paper using a new set of methods and additional data. The number of continuum-selected galaxies in each field is given in Table~\ref{table:fields} under the ``$N_\textrm{KBSS}$'' column. 

\begin{figure}
    \centering
    \includegraphics[width=0.47\textwidth]{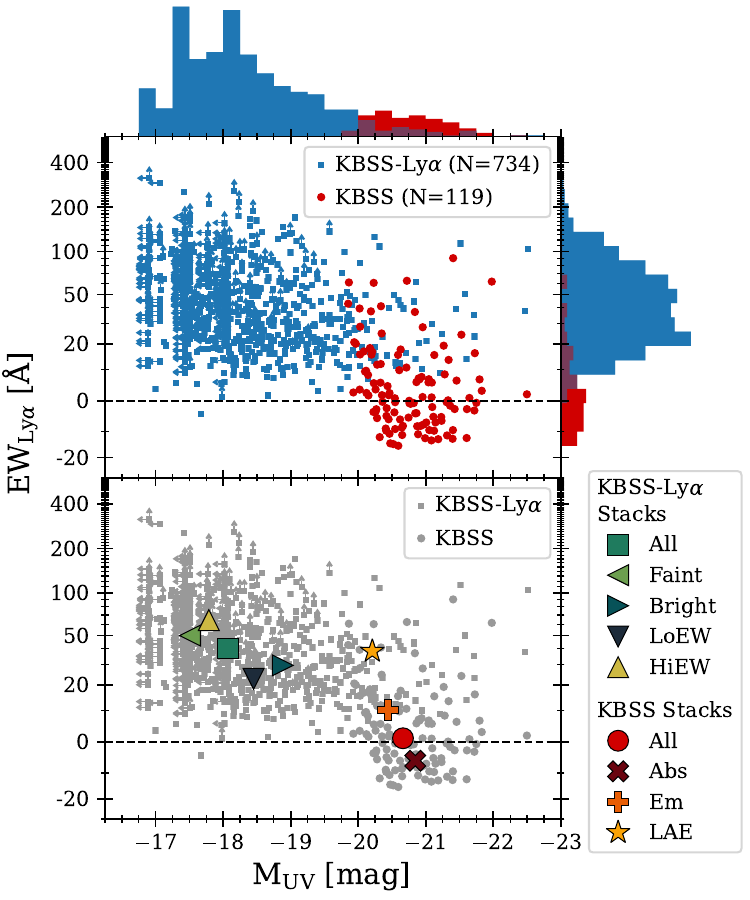}
    \caption{The distributions of KBSS (red) and KBSS-\lya\ (blue) galaxies with respect to EW$_{\mathrm{Ly}\alpha}$ and $M_\text{UV}$ as defined in the text. Top panel and histograms show the distributions of individual galaxy properties, while the bottom panel shows the median properties of each galaxy subsample described in Sec.~\ref{sec:subsamples} and Table~\ref{table:stacks}. The EW$_{\mathrm{Ly}\alpha}$ values are very sensitive to the choice of aperture; see note in Table~\ref{table:stacks} and additional discussion in Sec.~\ref{sec:curveofgrowth}.}
    \label{fig:cand_distributions}
\end{figure}

The distribution of \lya\ and UV continuum emission properties of the KBSS-\lya\ and KBSS galaxy samples are presented in Fig.~\ref{fig:cand_distributions} and Table~\ref{table:stacks}. As displayed in Fig.~\ref{fig:cand_distributions}, most of the continuum-bright and low-EW$_{\lya}$ objects described by our analysis are drawn from the KBSS---this sample is thus important for exploring the variation of \lya\ halos with galaxy properties.

\section{Analysis}
\label{sec:analysis}

\subsection{Creation of Ly$\alpha$ and continuum images} %name?
\label{sec:stamps}

In principle, Eq.~\ref{eq:flya}~\&~\ref{eq:fcont} could be applied directly to each pixel of our full-field NB and BB images to construct \lya\ and continuum images of each field. However, as noted in Sec.~\ref{sec:imaging}, there are small spatially-varying astrometric offsets between our images that complicate the spatially-resolved continuum subtraction and construction of the \lya\ and continuum images. With this in mind, we apply Eqs.~\ref{eq:flya}~\&~\ref{eq:fcont} to each galaxy individually after applying small astrometric corrections.

 We first cut out a square ``postage stamp'' centered on each galaxy. To optimize the astrometric alignment, we oversample each full-field image by a factor of two using the Scipy function \texttt{interpolate.RectBivariateSpline}. Square stamps with side length of 200 pixels (21$''$) are then extracted using the AstroPy function  \texttt{nddata.Cutout2D} with \texttt{mode='partial'}, which masks pixels that would fall outside the full-field image. Galaxies within 10 pixels ($\sim$1$''$) of the field edge are omitted from the sample. A 2D cross-correlation is then performed between the NB and BB stamps corresponding to each object. Because each stamp is typically large enough to detect multiple  foreground or background sources that should be identical in both images, the peak of the cross-correlation signal corresponds to the astrometric offset between the NB and BB stamps. Approximately half of the objects receive no shift, while 174 (63, 22, 11, 5, 2) objects are shifted by $>$1 (2, 3, 4, 5, 6) pixels, with a maximum 2D shift of  6.3 pixels (0\farcs66).  Eqs.~\ref{eq:flya}~\&~\ref{eq:fcont} are applied to the astrometrically-corrected stamps to construct individual \lya\ and continuum stamps for each galaxy.

\subsection{Galaxy subsamples}
\label{sec:subsamples}
Because previous work has suggested that \lya\ halo profiles may vary with galaxy properties such as continuum luminosity, environment, and EW$_{\lya}$, we divide our galaxies into several subsamples described in Table~\ref{table:stacks}.
In general, we consider the \lya-selected and continuum-selected galaxy samples separately, dividing each into subsets based on EW$_{\lya}$ and $M_\textrm{UV}$ measured from the NB, BB, and $\mathcal{R}_s$ images. In general, the faintness of the \lya-selected galaxies requires using large subsamples ($N_\text{gal}\gtrsim100$) in order to obtain sufficient signal to measure the large-scale halos; for the more luminous\footnote{Note that the continuum-selected galaxies are $\sim$10$\times$ brighter than the \lya-selected galaxies in the stellar continuum, but they have a wide range of \lya\ luminosities.} continuum-selected galaxies, samples of $N_\text{gal}\gtrsim20-50$ are sufficient depending on EW$_{\lya}$.

\subsection{Creation of galaxy stamps and stacks}
\label{sec:stacks}

For each galaxy subsample described in Sec.~\ref{sec:subsamples}, we average the 
\lya\ and continuum stamps for that sample to create 2D stacked \lya\ and continuum images. Each galaxy stamp is scaled based on its flux within a 1\farcs2 diameter aperture, and the set of stamps in that subsample is then sigma-clipped at each pixel location and averaged, weighted by these central fluxes. The scaling is performed in order to prevent individual bright sources from being censored by the sigma-clipping process. Galaxies with a central flux less than the median are scaled based on the median flux, and no scaling is performed for KBSS ``Abs'' or ``All'' \lya\ images for which the median flux is negative or nearly zero. While we experimented with masking background objects, we found that the sigma-clipping algorithm was sufficient to eliminate contamination from individual nearby sources, given variation in the sigma-clipping parameters with the properties of the sub-sample. The KBSS-\lya\ samples use a relatively low clipping threshold of 2.5$\sigma$ owing to the fact that even faint nearby sources produce significant contributions to the inferred background. Given the large number of galaxies combined for each of the KBSS-\lya\ stacks, this low threshold still allows us to average over many galaxies at each pixel position. Conversely, the KBSS galaxies are significantly brighter and have fewer galaxies per stack, so we use a higher clipping threshold of 4$\sigma$ for these stacks to avoid unnecessarily omitting too many sources from the average at each pixel.

\begin{figure*}
    \centering
    \includegraphics[width=\textwidth]{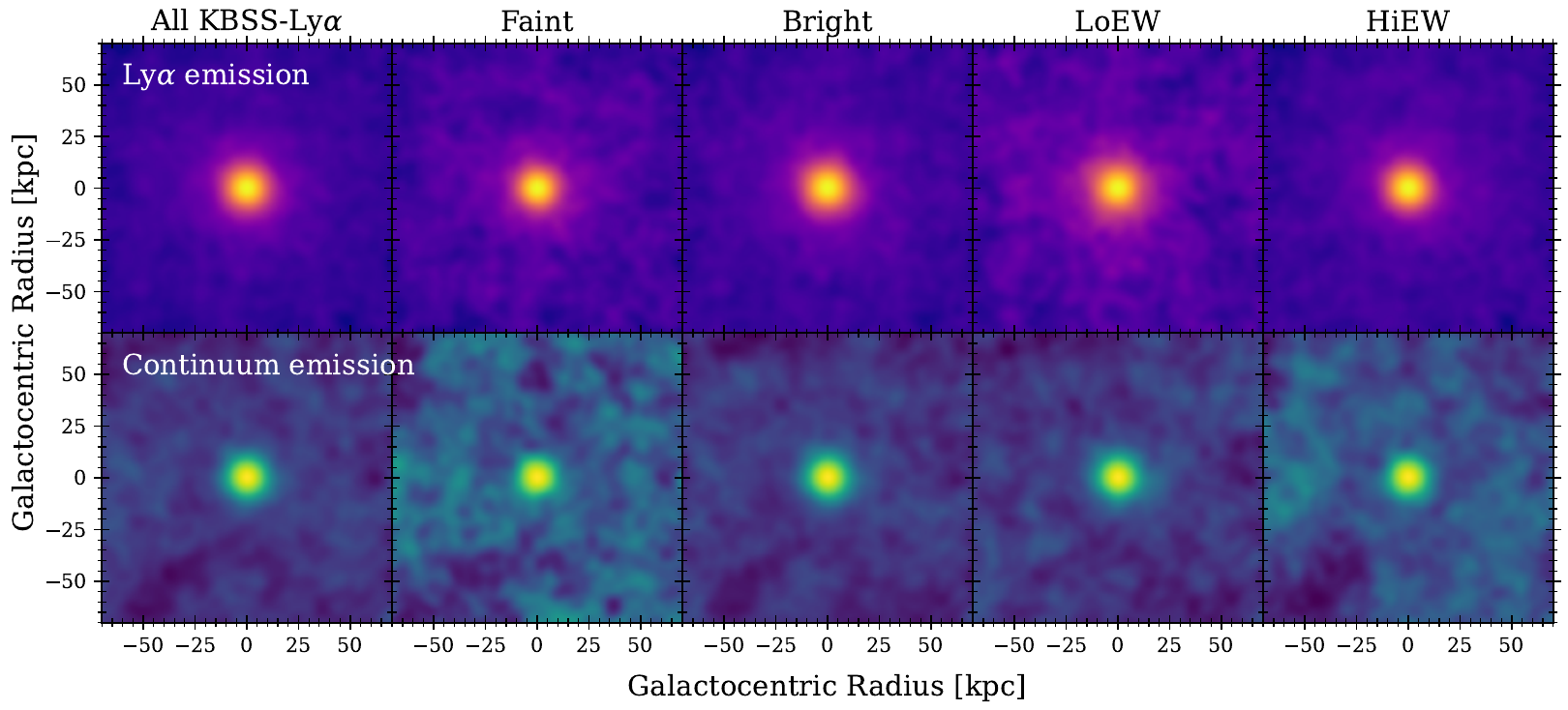}\vspace{4mm}
    
    \caption{Stacked \lya\ (top) and continuum (bottom) emission profiles for the KBSS-\lya\ (i.e., \lya-selected) subsamples. Objects are binned by continuum luminosity and EW$_{\textrm{Ly}\alpha}$ as described in Table~\ref{table:stacks}. Both sets of images are scaled by an arcsinh function, smoothed with a $\sigma=3\,\textrm{pixel}\approx0.3''$ gaussian kernel, and stretched to emphasize low surface-brightness emission.}
    \label{fig:2DPanel_kbsslya}
\end{figure*}

\begin{figure*}
    \centering
    \includegraphics[width=0.815\textwidth]{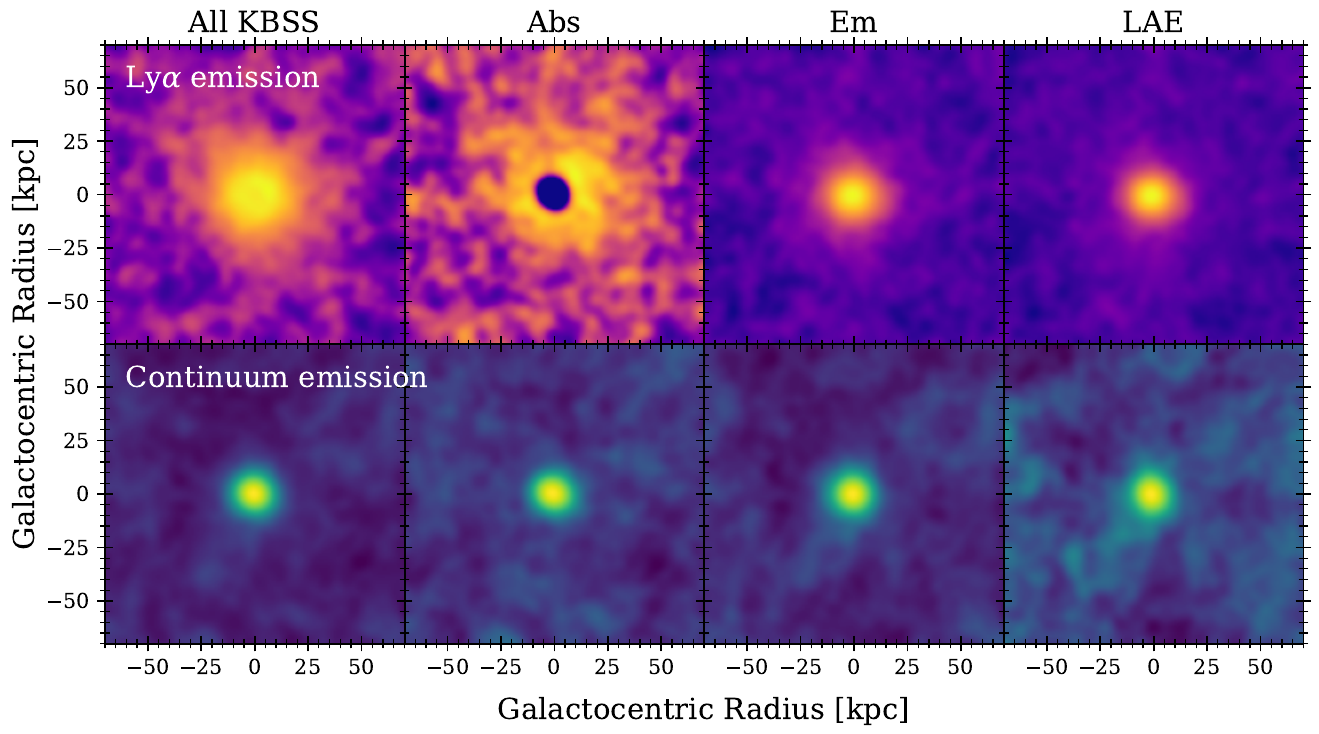}
    \caption{Stacked \lya\ (top) and continuum (bottom) emission profiles as in Fig.~\ref{fig:2DPanel_kbsslya} for the KBSS (i.e., continuum-selected) galaxy samples. Objects are binned by central \lya\ emission/absorption and EW$_{\textrm{Ly}\alpha}$ as described in Table~\ref{table:stacks}.}
    \label{fig:2DPanel_kbss}
\end{figure*}

The increase in S/N achieved by stacking the individual galaxy stamps in many cases revealed offsets in the sky-subtraction of our fields, with the continuum images being systematically over-subtracted near the central source such that the sky background appears to grow with radius. We therefore fit a linear radial background term to each of the stacked continuum images over the range $3.1''<\theta<10''$ ($25<r/\textrm{kpc}<80$) that is then subtracted from each continuum image. The \lya\ images do not display this radially-dependent oversubtraction, so a constant background term is subtracted from the \lya\ stacks based on the median pixel value at $8''<\theta<11''$ ($65<r/\textrm{kpc}<90$). Stacked Lya and continuum images for each galaxy subsample are displayed in Figs.~\ref{fig:2DPanel_kbss}~\&~\ref{fig:2DPanel_kbsslya}. 

For each stack, we also construct a set of bootstrap resampled stacks for uncertainty estimates. For each stack of $N$ galaxies, 100 bootstrap samples of $N$ galaxy images are selected (with replacement) and stacked according to the same methods described above, including sigma-clipping and background subtraction.  The standard deviation of values at each pixel location among the 100 bootstrap samples is used to estimate the flux uncertainty in 2D, and we also repeat many of the analyses in Sec.~\ref{sec:halos} on each of the bootstrap samples in order to infer uncertainties on derived quantities. 

For each stack, we also construct a one-dimensional, azimuthally-averaged flux distribution as shown in Figs.~\ref{fig:1Dlya_kbsslya}--\ref{fig:1Dcont_kbss}. Annular averages are calculated as the (3$\sigma$) sigma-clipped mean of the pixels within the corresponding annulus. The flux uncertainty is the standard error of that mean as calculated from the standard deviation of the included pixels divided by $\sqrt{N}/2$, where the factor of 2 accounts for the correlations among pixels in our oversampled images.

\section{Halo Measurements}
\label{sec:halos}
\subsection{Direct exponential fits to 1D profiles}
\label{sec:expfits}

For consistency with the \lya\ halo measurements by \citet{steidel2011}, our initial estimates of the galaxy light profiles are derived from direct fits of exponential profiles to the empirical azimuthally-averaged distributions of \lya\ and continuum emission. The fit profiles take the following form:
\begin{equation}
S(r)=S_0\,e^{-r/r_{0,\mathrm{dir}}}
\label{eq:exp}
\end{equation}
\noindent where $S(r)$ is the azimuthally-averaged surface brightness at a given galactocentric projected distance $r$, $S_0$ is the amplitude, and $r_{0,\mathrm{dir}}$ is the exponential scale length.

\begin{figure*}
\includegraphics[width=\textwidth]{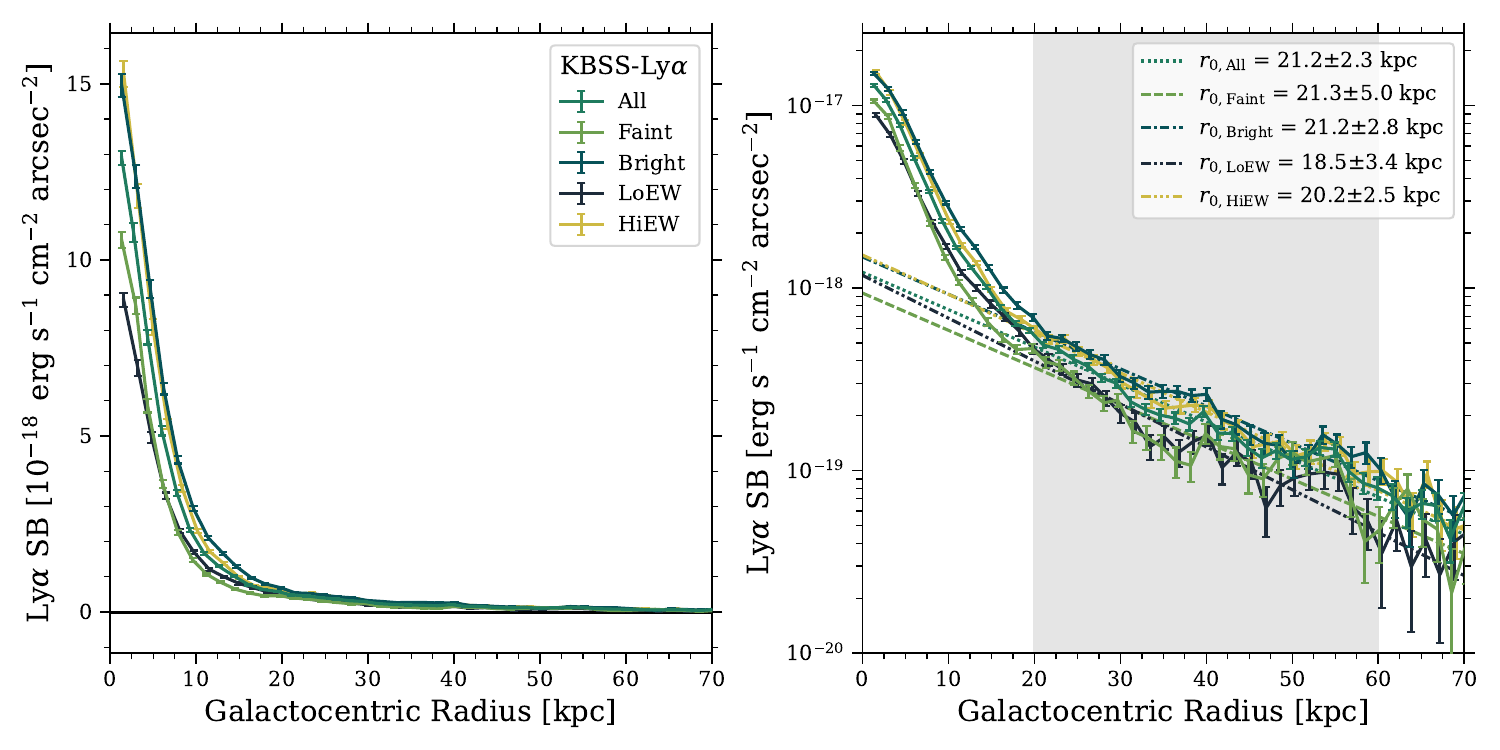}
\caption{Azimuthally-averaged \lya\ surface brightness profiles for stacked galaxy samples drawn from the KBSS-\lya. Both panels show the same data with linear (left) or logarithmic (right) scaling in \lya\ surface brightness. Colored lines correspond to the galaxy stacks named in the legend of the left panel; detailed information for each subsample is given in Table~\ref{table:stacks} and in Sec.~\ref{sec:stacks}. The shaded region in the right panel shows the range of galactocentric projected distances over which exponential profiles are fit, with the resulting best-fit profiles displayed as dashed or dot-dashed lines and the corresponding exponential scale lengths included in the legend. In general, the \lya\ profiles are highly consistent among these subsamples.}\label{fig:1Dlya_kbsslya}
\end{figure*}

\begin{figure*}
\includegraphics[width=\textwidth]{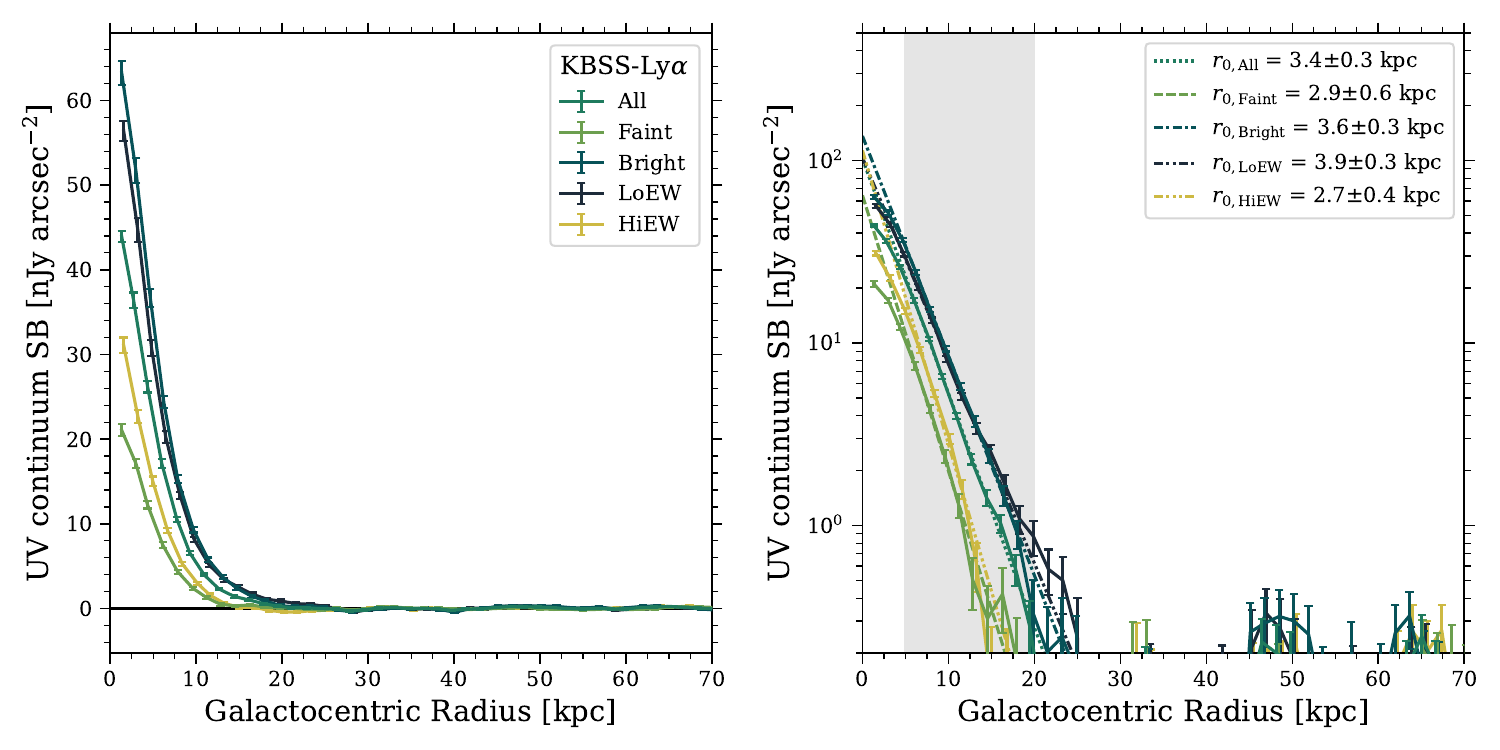}
\caption{Azimuthally-averaged continuum surface brightness profiles for stacked galaxy samples drawn from the KBSS-\lya, which are analogous to the \lya\ profiles in Fig.~\ref{fig:1Dcont_kbsslya}. As above, only the scaling differs between the two panels, and shading in the right panel denotes the region over which the exponential profiles are fit. As for the \lya\ profiles, the continuum emission profiles are largely consistent among the KBSS-\lya\ subsamples.}\label{fig:1Dcont_kbsslya}
\end{figure*}

\begin{figure*}
\includegraphics[width=\textwidth]{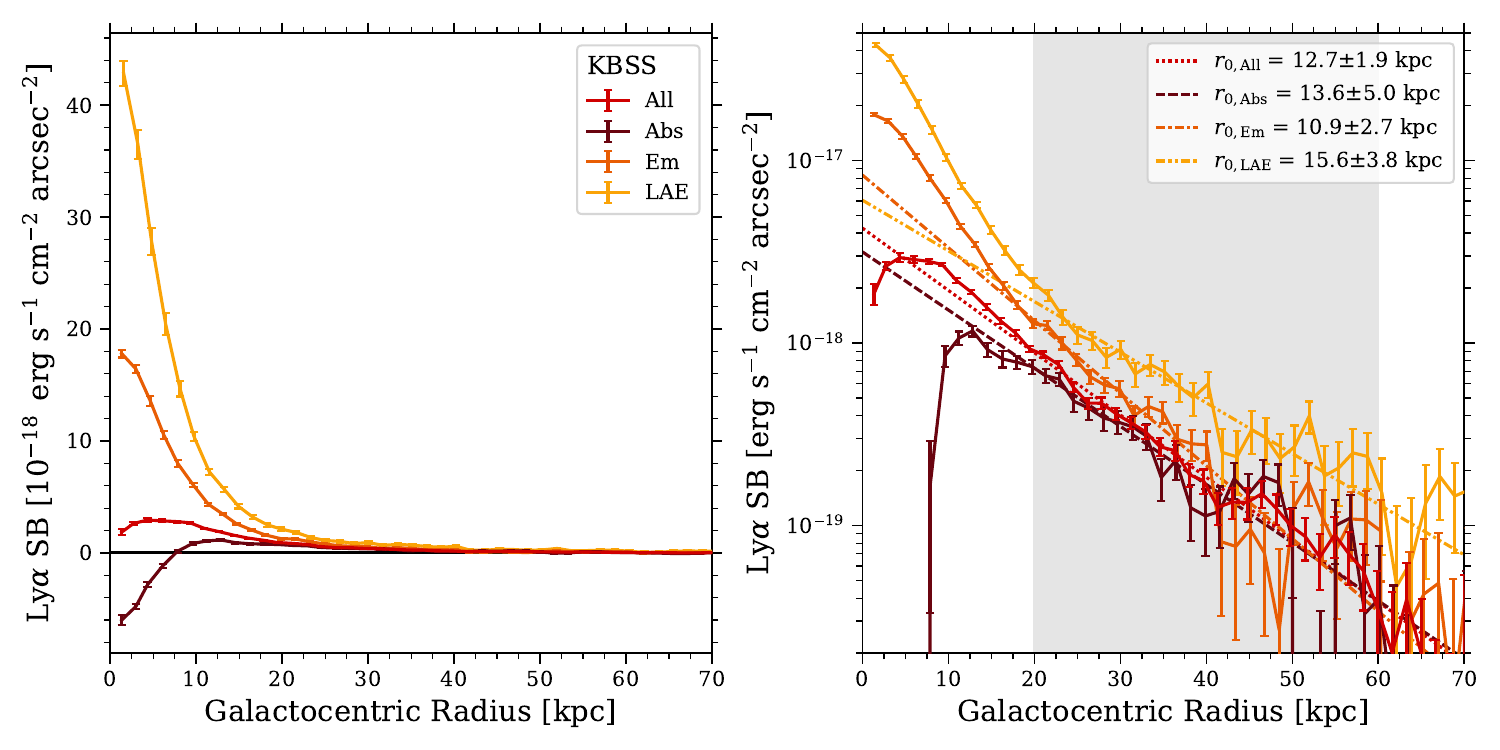}
\caption{Azimuthally-averaged \lya\ surface brightness profiles as in Fig.~\ref{fig:1Dlya_kbsslya} for stacks of continuum-selected galaxies from the KBSS. Samples are selected based on their central \lya\ emission as described in Table~\ref{table:stacks}. Because the KBSS galaxies have a broad range of central \lya\ emission properties (including net \lya\ absorption at small projected distances for many galaxies), the shapes of the inner profiles vary much more significantly than those in Fig.~\ref{fig:1Dlya_kbsslya}. However, the large-scale \lya\ profile morphologies and exponential scale lengths are highly consistent among the stacked KBSS subsamples.}\label{fig:1Dlya_kbss}
\end{figure*}

\begin{figure*}
\includegraphics[width=\textwidth]{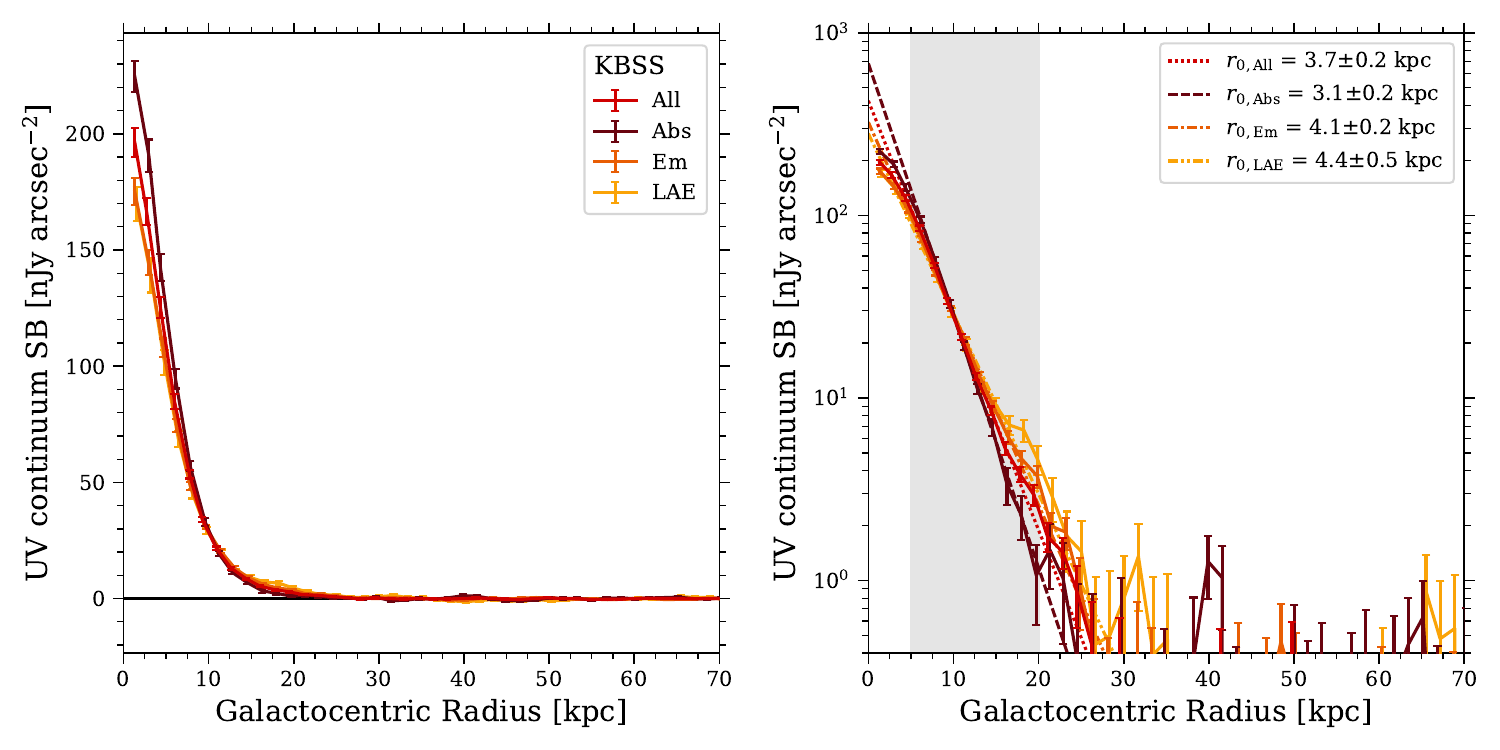}
\caption{Azimuthally-averaged continuum surface brightness profiles as in Fig.~\ref{fig:1Dcont_kbsslya} for stacks of continuum-selected galaxies from the KBSS. Again, the continuum emission is highly consistent among the stacked subsamples.}\label{fig:1Dcont_kbss}
\end{figure*}

Figs.~\ref{fig:1Dlya_kbsslya}~\&~\ref{fig:1Dlya_kbss} display the fitted \lya\ halo profiles for stacked subsamples of galaxies drawn from the KBSS-\lya\ and KBSS, respectively. The grey shaded regions show the range of projected radii over which the profiles are fit to the data (20 kpc $<r<$ 60 kpc), approximately the range over which the extended \lya\ component is detected at high significance for all of the stacks. Each annulus within the fitted region is weighted by its corresponding uncertainty, and the best-fit scale lengths are given in the figure legends. Similar fits to the continuum flux distribution are given in  Figs.~\ref{fig:1Dcont_kbsslya}~\&~\ref{fig:1Dcont_kbss}, where again the shaded region identifies the range of annuli over which the exponential models are fit (5 kpc $<r<$ 20 kpc).

Uncertainties on the exponential scale lengths are derived by performing the profile-fitting method above on each of the 100 bootstrap stacks described in Sec.~\ref{sec:stacks}, and the standard deviation of the scale lengths from these 100 fits is reported as the 1$\sigma$ uncertainty on that value in Table~\ref{table:fits} and in Figs.~\ref{fig:1Dlya_kbsslya}$-$\ref{fig:1Dcont_kbss}.

The \lya\ profiles are significantly extended with respect to the continuum emission in a manner qualitatively similar to previous studies. The scale lengths for each stack are broadly similar, with typical values $r_{0,\mathrm{dir}}\approx 10-20$~kpc for the \lya\ emission and $r_{0,\mathrm{dir}}\approx 3$~kpc in the continuum. Notably, the continuum scale lengths are similar to those derived by \citet{steidel2011} using similar techniques, but the derived \lya\ scale lengths in that paper are somewhat larger ($r_0^\textrm{Ly$\alpha$}\approx20-30$ kpc). We discuss the similarities and differences among these and other recent \lya\ halo measurements in Sec.~\ref{sec:discussion}.

Note that the derived exponential scale lengths depend on the range of annuli over which they are fitted. For the \lya\ profiles, the fit region is chosen to be that for which the profile is monotonic and significantly detected for the majority of the stacks, which occurs at $r\gtrsim 20$ kpc; Fig.~\ref{fig:1Dlya_kbss} clearly shows that the behavior at smaller projected distances varies dramatically among the KBSS stacks, and nearly all the KBSS and KBSS-\lya\ stacks show a change in slope near this inner radius. Likewise, the fit region for the continuum stacks is chosen to be the range of radii with significant continuum emission, while avoiding the innermost annuli where the profiles visibly diverge from exponential behavior, presumably owing to the seeing-limited point-spread function. Small changes to either range of fit annuli do not significantly change the fit scale lengths compared to the estimated uncertainties, so long as they do not include these innermost bins ($r\lesssim20$ kpc for \lya; $r\lesssim5$ kpc for the continuum).
The inferred scale lengths are also sensitive to the background subtraction, which is why we make a local background correction for each stack prior to fitting as described in Sec.~\ref{sec:stacks}.

The exponential fits above appear to approximate the large-scale behavior of the projected surface brightness distributions for \lya\ and UV continuum emission. However, the potential sensitivity of these fits to the modeled region, local background subtraction, and PSF of the observations motivates a more comprehensive modeling of their profiles over the full range of measured scales. We describe below a complementary forward-modeling approach to fitting the galaxy profiles that is also more comparable to other recent students of \lya\ halos (e.g., \citealt{leclercq2017}). A summary of the direct and forward-modeling fits is presented in Table~\ref{table:fits}.

\subsection{Forward modeling of halo profiles}
\label{sec:mcmcfits}

Our forward-modeling approach is designed to explicitly treat the central galaxy light profile, the halo profile, the sky background, and the PSF in a separable manner.

Specifically, we model the \lya\ and continuum surface brightness profiles using an azimuthally-symmetric model consisting of one or more exponential profiles convolved with the empirical PSF and added to a uniform background. Although the surface brightness models are symmetric (and thus 1D), we fit them directly to the 2D image constructed for each stacked galaxy sample.

\begin{figure*}
\includegraphics[width=\textwidth]{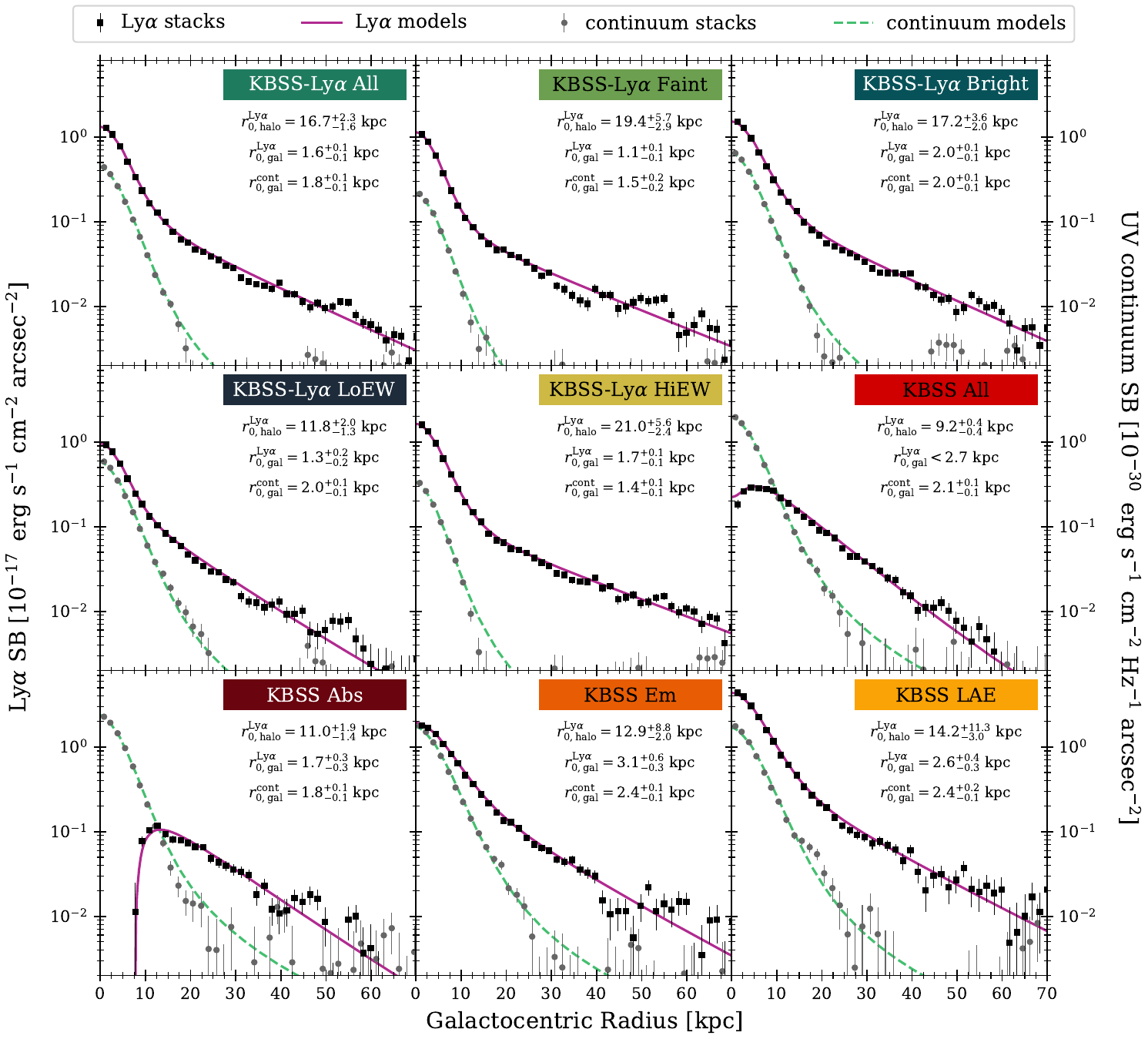}
\caption{Azimuthally-averaged \lya\ and continuum surface brightness profiles and fit results from our forward-modeling framework, which incorporates the large-scale Keck/LRIS-B PSF and uncertainties in the local background subtraction (Sec.~\ref{sec:mcmcfits}). The empirical \lya\ (continuum) distribution is displayed as black squares (gray circles). The solid purple and dashed teal lines represent the corresponding best-fit models. Both the models and the data reflect the subtraction of the best-fit estimate of the uniform local sky background, so the surface brightness at each projected position generally varies slightly with respect to those displayed in Figs.~\ref{fig:1Dlya_kbsslya}--\ref{fig:1Dcont_kbss} where the sky background is estimated independently from the galaxy model.}
\label{fig:mcmc_plot_grid}
\end{figure*}

We fit the \lya\ profile with two exponential profiles to account for an extended ``halo''  in addition to a central ``galactic'' component.\footnote{This dual-exponential model is used to provide consistency with other recent work (e.g., \citealt{leclercq2017}) and because it provides a good qualitative description of the empirical profiles over a range of galaxy properties; see Fig.~\ref{fig:mcmc_plot_grid}.} The halo term is omitted while fitting the continuum emission, although the amplitude of the best-fit halo is generally found to be consistent with zero even when it is included in the continuum fit. The complete expression for the \lya\ and continuum models are thus:
\begin{align}
S_\mathrm{Ly\alpha}(r)&=\left[S^\mathrm{gal}_\mathrm{exp}(r)+S^\mathrm{halo}_\mathrm{exp}(r)\right]*S_\mathrm{PSF}(r)+S_\mathrm{bg}\label{eq:slya}\\
S_\mathrm{cont}(r)&=S^\mathrm{gal}_\mathrm{exp}(r)*S_\mathrm{PSF}(r)+S_\mathrm{bg}\label{eq:scont}
\end{align}
\noindent where $*$ denotes the PSF convolution, and the exponential terms are defined as below:
\begin{align}
S_\mathrm{exp}(r)=\frac{A_R\,e^{-r/r_0}}{C}
\end{align}

As shown above, each exponential has two independent parameters, the scale length $r_\mathrm{0}$ and the total flux $A_R$ within a fixed radius $R$. We choose $R=50\,\textrm{kpc}\approx6''$ for reasons discussed below. $C$ is a normalization factor that is determined by $R$ and $r_0$ such that $S_\mathrm{exp}(r=0)=A_R/C$. For a given $r_0$ and $R$, $C$ is thus defined by the following expression:
\begin{equation}
%C=2 p^2 \pi r_\mathrm{s}
C\equiv2 \pi r_0\left(r_0-(r_0+R)e^{-R/r_0}\right)\quad .
\end{equation}
We choose to parameterize our model in terms of $A_{R}=A_{50}$ rather than in terms of the central surface brightness $S_\mathrm{exp}(r=0)$ because the integrated flux enclosed by the \lya\ halo is more directly constrained by our data than the amplitude of the halo term at $r=0$, the latter parameter being strongly covariant with the scale length $r_0$.  The choice of $R=50\,$kpc is motivated in part because it is approximately the radius at which the enclosed flux exhibits the least covariance with respect to $r_0$ for our data. In addition, 50 kpc is the approximate virial radius $r_\textrm{vir}=(3M_\textrm{DM}/800\pi \rho_\textrm{crit})^{1/3}$ for the galaxies in our sample; at $z=2.6$, our galaxies with $M_{*}\approx 10^{8}-10^{11}\,$M$_\odot$ likely inhabit dark matter halos with  $M_\textrm{DM}\approx 10^{11}-10^{12}\,$M$_\odot$\footnote{This range is based on the mean stellar-to-halo mass relationship for $z\approx2-3$ estimated by \citet{beh13}, and \citet{trainor2012} also find $M_\textrm{DM}\approx 8\times10^{11}\,$M$_\odot$ from a clustering analysis of the KBSS galaxies in this sample.}, corresponding to $r_\textrm{vir}\approx40-80\,$kpc. Thus our measurements of the \lya\ profile within a radius $R\approx 50\,$kpc approximates the emission within the CGM.

Note that $A_{50}$ is defined separately for each exponential component of the \lya\ fit and may be positive or negative: a negative $A_\textrm{50}$ occurs for the inner ``galaxy'' term for several stacks that exhibit central \lya\ absorption surrounded by extended emission. Note that while our two-exponential model is motivated by the similar models used by \citet{wisotzki2016} and \citet{leclercq2017} to fit the \lya\ profiles of \lya-emitters, to the best of our knowledge this model has not been previously used to self-consistently fit profiles exhibiting central \lya\ emission as well as \lya\ absorption. This framework thus represents a novel aspect of our study, and---as demonstrated below---it provides both a successful fit to our data as well as interpretative power for understanding the nature of central \lya\ absorption.

\begin{deluxetable*}{l@{\hspace{1cm}}cc@{\hspace{1cm}}cccc}%{lcrcrcr@{}l}
    \renewcommand{\arraystretch}{1.1}
    \tablecaption{Best-Fit Model Parameters}
    \tablewidth{0pt}
    \tablehead{
    % \colhead{QSO Field} & $z\sub{Q}$ ($\pm$0.001) & NB Filter & FWHM (\AA)
    % & Exp (s) & BB Filter & $N_\textrm{KBSS}$ & $N_\textrm{KBSS-Ly$\alpha$}$
    % \multicolumn{2}{c}{NB Filter} & \colhead{Sample}  & FWHM (\AA)
    % & Exp (s) & BB Filter & $N_\textrm{KBSS}$ & $N_\textrm{KBSS-Ly$\alpha$}$
    Sample &
    $r_{0,\textrm{dir}}^\textrm{Ly$\alpha$}$ (kpc)&
    $r_{0,\textrm{dir}}^\textrm{cont}$ (kpc)&
    $r_{0,\textrm{halo}}^\textrm{Ly$\alpha$}$ (kpc)&
    $r_{0,\textrm{gal}}^\textrm{Ly$\alpha$}$ (kpc)&
    $r_{0,\textrm{gal}}^\textrm{cont}$ (kpc)&
    $f_\textrm{gal}^\textrm{Ly$\alpha$}$ (\%)
    }\colnumbers
    \startdata
    \multicolumn{4}{l}{\it KBSS-\lya\ galaxy samples}\\
    All   & 21.2$\pm$2.3 &  3.4$\pm$0.3 & 16.7$_{- 1.6}^{+ 2.3}$ &  1.6$_{- 0.1}^{+ 0.1}$ &  1.8$_{- 0.1}^{+ 0.1}$ & \phantom{{+}} 44$_{- 1}^{+ 1}$\\
    Faint & 21.3$\pm$5.0 &  2.9$\pm$0.6 & 19.4$_{- 2.9}^{+ 5.7}$ &  1.1$_{- 0.1}^{+ 0.1}$ &  1.5$_{- 0.2}^{+ 0.2}$ & \phantom{{+}} 41$_{- 2}^{+ 3}$\\
    Bright & 21.2$\pm$2.8 &  3.6$\pm$0.3 & 17.2$_{- 2.0}^{+ 3.6}$ &  2.0$_{- 0.1}^{+ 0.1}$ &  2.0$_{- 0.1}^{+ 0.1}$ & \phantom{{+}} 47$_{- 2}^{+ 2}$\\
    LoEW & 18.5$\pm$3.4 &  3.9$\pm$0.3 & 11.8$_{- 1.3}^{+ 2.0}$ &  1.3$_{- 0.2}^{+ 0.2}$ &  2.0$_{- 0.1}^{+ 0.1}$ & \phantom{{+}} 36$_{- 3}^{+ 2}$\\
    HiEW & 20.2$\pm$2.5 &  2.7$\pm$0.4 & 21.0$_{- 2.4}^{+ 5.6}$ &  1.7$_{- 0.1}^{+ 0.1}$ &  1.4$_{- 0.1}^{+ 0.1}$ & \phantom{{+}} 47$_{- 2}^{+ 3}$\\[5pt]\hline
    \multicolumn{4}{l}{\it KBSS galaxy samples}\\
    All &  12.7$\pm$1.9 &  3.7$\pm$0.2 &   9.2$_{- 0.4}^{+ 0.4}$ &  $<2.7$ &  2.1$_{- 0.1}^{+ 0.1}$ & \phantom{1}$-$7$_{- 2}^{+ 2}$\\
    Abs &  13.6$\pm$5.0 &  3.1$\pm$0.2 &  11.0$_{- 1.4}^{+ 1.9}$ &  1.7$_{- 0.3}^{+ 0.3}$ &  1.8$_{- 0.1}^{+ 0.1}$ & \phantom{$^1$}$-$65$_{-14}^{+18}$\\
    Em & 10.9$\pm$2.7 &  4.1$\pm$0.2 & 12.9$_{- 2.0}^{+ 8.8}$ &  3.1$_{- 0.3}^{+ 0.6}$ &  2.4$_{- 0.1}^{+ 0.1}$ & \phantom{{+}} 48$_{- 7}^{+ 5}$\\
    LAE &  15.6$\pm$3.8 &  4.4$\pm$0.5 &  \phantom{$^1$}14.2$_{- 3.0}^{+11.3}$ &  2.6$_{- 0.3}^{+ 0.4}$ &  2.4$_{- 0.1}^{+ 0.2}$ & \phantom{{+}} 57$_{- 4}^{+ 5}$\\
    \enddata
    \tablecomments{Columns (2)--(3) correspond to parameters of the direct exponential fits (Sec.~\ref{sec:expfits}), with $r_{0,\textrm{dir}}$ defined by Eq.~\ref{eq:exp}. Columns (4)--(7) correspond to parameters of the forward-modeling technique (Sec.~\ref{sec:mcmcfits}), with parameters defined by Eqs.~\ref{eq:slya}--\ref{eq:fgal}. The value of $r_{0,\textrm{gal}}^\textrm{Ly$\alpha$}$ for the KBSS-All sample is given as a 2$\sigma$ upper limit as discussed in Sec.~\ref{sec:centralcomparison}.}
    \label{table:fits}
\end{deluxetable*}

Each Ly$\alpha$ stack is fit using a seven-parameter model: the x- and y- coordinates of the center of the galaxy; the scale length $r_0$ and enclosed flux $A_{50}$ for each of the two exponential terms, and the constant background. 
The continuum stacks are modeled using only one exponential model representing the galaxy (i.e., they are not fit using a halo term) but are otherwise identical to the \lya\ model, so the continuum fits include five free parameters. As we are primarily interested in understanding the shape of the \lya\ profile with respect to the continuum profile, our primary parameters of interest are the scale radii of the \lya\ halo ($r_{0,\textrm{halo}}^\textrm{Ly$\alpha$}$), the central \lya\ component ($r_{0,\textrm{gal}}^\textrm{Ly$\alpha$}$), and the continuum emission ($r_{0,\textrm{gal}}^\textrm{cont}$). We also define a parameter that describes the relative \lya\ luminosity of the ``galaxy'' and ``halo'' components:
\begin{equation}
f_\textrm{gal}^\textrm{Ly$\alpha$}=\frac{A_\textrm{50,gal}}{A_\textrm{50,gal}+A_\textrm{50,halo}}
\label{eq:fgal}
\end{equation}
\noindent such that $f_\textrm{gal}^\textrm{Ly$\alpha$}=1$  describes a galaxy with no halo component to the \lya\ surface brightness distribution, and $f_\textrm{gal}^\textrm{Ly$\alpha$}=0$ would describe a galaxy with a pure \lya\ halo with no central component.\footnote{Note that we require $r_{0,\textrm{gal}}^\textrm{Ly$\alpha$}<5$ kpc and $r_{0,\textrm{halo}}^\textrm{Ly$\alpha$}>5$ kpc to distinguish between these two cases.} For galaxies exhibiting central \lya\ absorption with surrounded by a \lya-emitting halo, both $A_\textrm{50,gal}$ and $f_\textrm{gal}^\textrm{Ly$\alpha$}$ are negative.

An empirical PSF is constructed for each field from light profiles of point sources in the Keck/LRIS-B images. Stars over a wide range of magnitudes ($16\lesssim B \lesssim 21$) are combined to determine the PSF over the range $0\le\theta\lesssim10''$, thus extending well beyond the range of angular scales fit by our profile measurements. Because our halo measurements combine galaxy samples across multiple fields, an effective PSF is constructed by averaging the field-specific PSFs. A detailed description of the PSF construction is given in Appendix~\ref{sec:psf}. The constructed empirical PSF is sampled at the same spatial scale as the images (i.e., twice the native resolution of our \lya\ and continuum images) and numerically convolved with the analytic function describing the underlying flux distributions (Eqs.~\ref{eq:flya}~\&~\ref{eq:fcont}) using the Python function \texttt{scipy.signal.fftconvolve}.

Fitting is conducted using the Markov-Chain Monte Carlo (MCMC) modeling library \texttt{emcee} \citep{foreman-mackey2013}. Flat prior distributions are assumed for each parameter except for the background term $S_\textrm{bg}$, which has a gaussian prior with zero mean\footnote{Note that the images input to the forward-modeling fit have already had local background corrections applied as described in Sec.~\ref{sec:stacks}, so the background term in the model reflects the uncertainty in this correction.} and $\sigma$ given by the standard deviation of the backgrounds estimated among the 100 bootstrap stacks for that sample (Sec.~\ref{sec:stacks}). Posterior distributions for each parameter are visually inspected for each of the fits to ensure convergence.

The best-fit \lya\ and continuum profiles obtained by our forward-modeling framework are given for each stacked galaxy subsample in Fig.~\ref{fig:mcmc_plot_grid}, with parameter values noted in that figure and/or Table~\ref{table:fits}. Comparisons among the direct fit and forward-modeling parameters discussed in Sec.~\ref{sec:comparison}.

% \subsubsection{Results of forward modeling}
% \label{sec:mcmcresults}

\subsection{\lya\ curve of growth}
\label{sec:curveofgrowth}

The extended nature of the \lya\ emission profile relative to the stellar continuum causes the value of EW$_{\textrm{Ly}\alpha}$ for a given galaxy to vary according the aperture used to measure these fluxes. This variation has significant effects for the identification of \lya\ in slit spectroscopy, such that only $\sim$50\% of $L\approx L_*$ continuum-selected galaxies at $z\approx2-3$ display net \lya\ emission in slit spectra (e.g., \citealt{shapley2003,steidel2010,trainor2019}). In general, the measured emission profile depends on the detailed \lya\ spatial profile and slit orientation, variation that can now be characterized using integral-field spectroscopy of individual galaxies (e.g., \citealt{wisotzki2016,leclercq2017,chen2021}). However, the aperture dependence of the averaged, azimuthally-symmetric \lya\ halo profiles described here can be described in terms of a one dimensional curve of growth: the integrated EW$_{\textrm{Ly}\alpha}$ within a circular aperture as a function of the aperture radius.

\begin{figure*}
\includegraphics[width=0.48\textwidth]{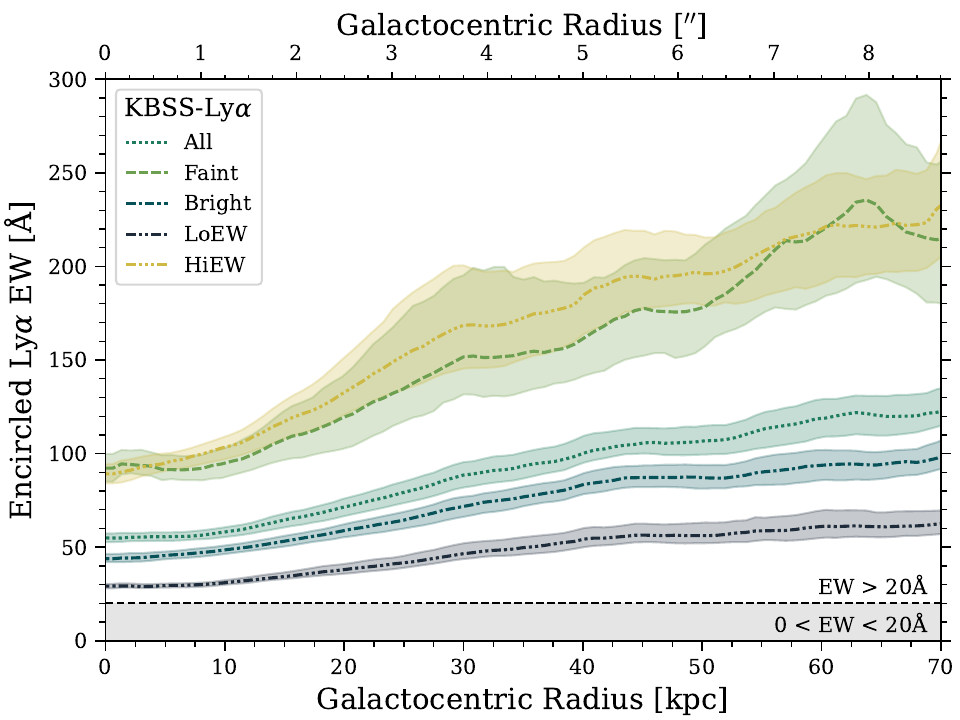}\hspace{0.04\textwidth}
\includegraphics[width=0.48\textwidth]{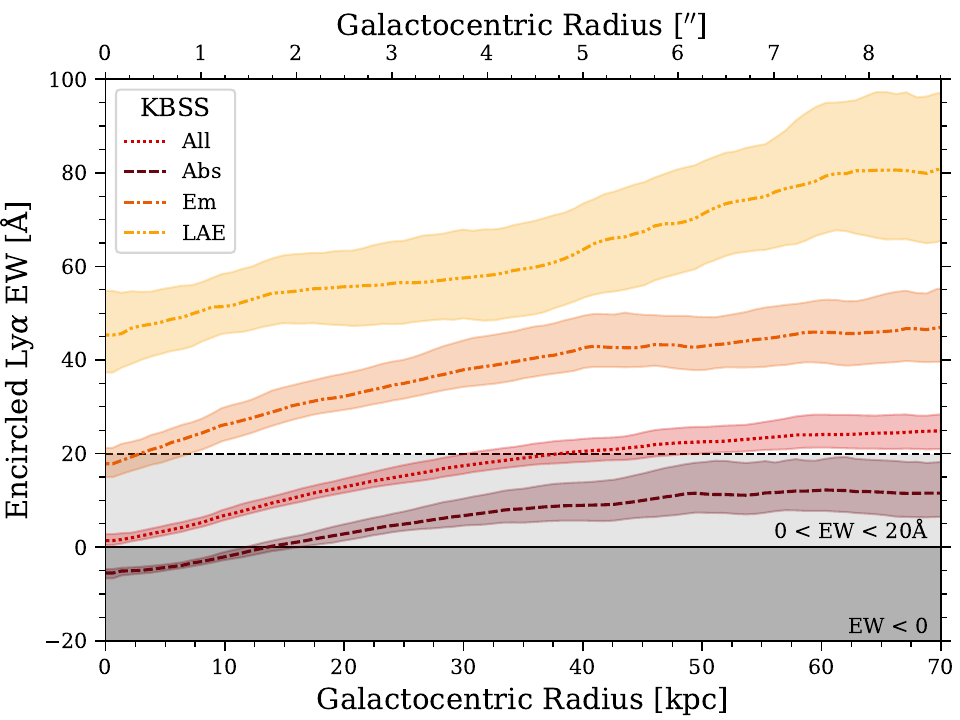}
\caption{Encircled EW$_{\textrm{Ly}\alpha}$ measured directly from the stacked galaxy samples presented in Figs.~\ref{fig:1Dlya_kbsslya}--\ref{fig:1Dcont_kbss}. Left panel displays the KBSS-\lya\ samples; right panel displays the corresponding curves for the KBSS samples. Shaded regions denote the 16\%-tile to 84\%-tile confidence interval derived from bootstrap stacks, and the dashed line denotes the 50\%-tile (median) curve for each sample. Horizontal lines bounding the gray shaded regions indicate  EW$_{\textrm{Ly}\alpha}=0$ and EW$_{\textrm{Ly}\alpha}=20$\AA, the latter being the typical cutoff for the photometric selection of \lya-emitters as in this paper. The EW$_{\textrm{Ly}\alpha}$ values grow with aperture radius in all cases because of the extended \lya\ emission; even those galaxies selected to be \lya\ absorbers at small radii are net \lya\ emitters within a sufficiently large aperture.}\label{fig:EWcog}
\end{figure*}

Fig.~\ref{fig:EWcog} displays the empirical encircled 
% EW$_{\textrm{Ly}\alpha}$ 
$\ewlya(<r)$
for each of the subsamples considered in this work, as well as the 1$\sigma$ uncertainty on $\ewlya(<r)$ at each radius derived from the bootstrap stacks described in Sec.~\ref{sec:stacks}. In each of the extracted curves, the growth of EW$_{\textrm{Ly}\alpha}$ with radius is apparent; again, this trend reflects the extended nature of the \lya\ emission with respect to the continuum. For the KBSS-\lya\ samples and and KBSS LAE stack, $\ewlya(<r)$ for $r\approx$~50$-$70 kpc is approximately 2$-$3$\times$ the value at $r=0$. The growth in EW$_{\textrm{Ly}\alpha}(<r)$ with radius is even more notable for the samples with low (or no net) central \lya\ emission: the KBSS Abs and KBSS All stacks have net absorption at $r=0$ but display positive $\ewlya(<r)$ at large radii, even approaching or exceeding the EW$_{\textrm{Ly}\alpha}>20$\AA\ threshold used for photometric classification of \lya-emitters. These measurements thus illustrate the lack of a consistent dichotomy between \lya-emitters and \lya-absorbers, reinforcing that these categories of galaxies must always be defined with respect to a particular aperture, as has also been highlighted in previous work (e.g., \citealt{steidel2011}).

Another interesting result from these curve-of-growth measurements are the similarities and differences in the large-scale behavior of the curves, most of which approximately parallel each other at large galactocentric radii. The consistency in the shapes of these profiles primarily reflects the similarities in the inferred halo scale lengths among all the samples. However, the vertical offsets among the $\ewlya(<r)$ curves are approximately constant with radius for the low-\ewlya\ samples, but grow approximately linearly with radius for the high-\ewlya\ samples. This qualitative variation in the relationships among the small-scale and large-scale \ewlya\ values is further explored in Sec.~\ref{sec:fgal}.

\section{Parameter Comparisons}
\label{sec:comparison}

\subsection{Galaxy-component \lya\ fraction}
\label{sec:fgal}

\begin{figure*}
\centering
\includegraphics[height=0.32\textwidth]{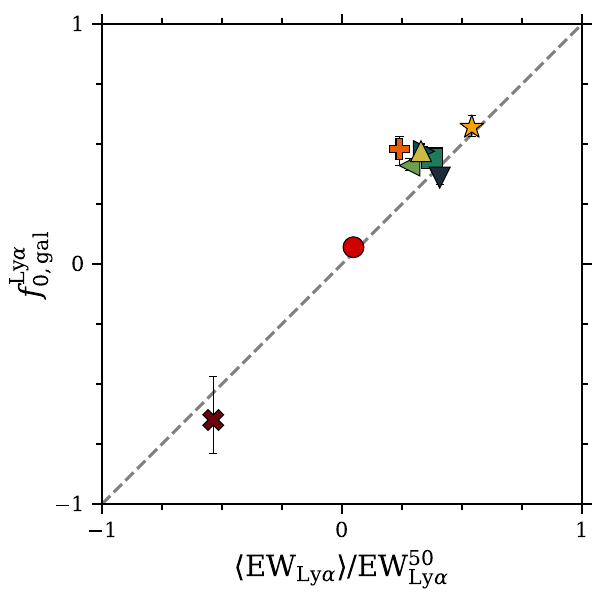}\hspace{4mm}
\includegraphics[height=0.32\textwidth]{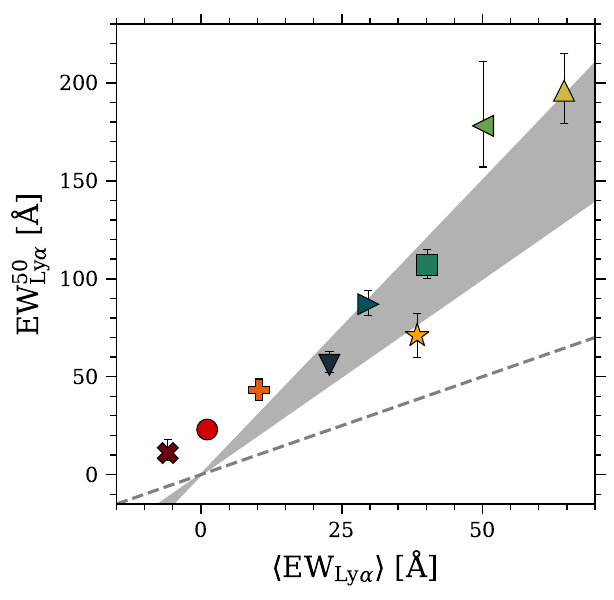}\hspace{4mm}
\includegraphics[height=0.32\textwidth]{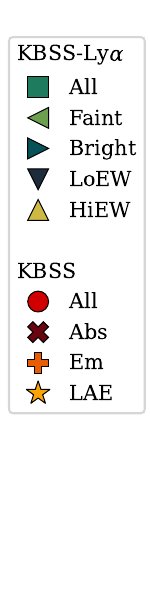}
\caption{
\textit{Left:} The fraction $f_{\textrm{gal}}^\textrm{Ly$\alpha$}$ of \lya\ flux associated with the central galaxy component as inferred from the forward-modeling fits (Sec.~\ref{sec:mcmcfits}) vs. the ratio between the median \ewlya\ value (calculated in a $\sim$5 kpc aperture) within a sample to the $\ewlya^{50}$ value measured on larger scales ($R=50$ kpc) from the stack of the same sample. $f_{\textrm{gal}}^\textrm{Ly$\alpha$}$ is almost identical to the direct ratio of these two \ewlya\ values. \textit{Right:} The two \ewlya\ values have a strong positive correlation among our samples, but the large-scale $\ewlya^{50}$ lies well above the dashed 1:1 line. The shaded region denotes $\ewlya^{50}=(2-3)\times\ewlya$, which is broadly consistent with the measurements of the high-\ewlya\ samples. However, the low-\ewlya\ samples have higher large-scale \ewlya\ values than those predicted by such a linear relationship, with a positive $\ewlya^{50}$ in all cases.}
\label{fig:fgal}
\end{figure*}

The parameter $f_{\textrm{gal}}^\textrm{Ly$\alpha$}$ in the forward-modeling fit describes the fraction of the galaxy's total \lya\ emission associated with the central galaxy component. This value describes the central concentration of the \lya\ emission (or absorption), and is thus closely related to the ratio of the \ewlya\ values calculated within the 1\farcs2 diameter aperture used to measure \ewlya\ for individual objects  and the $\ewlya^{50}$ values calculated from each stack within the larger ($R=50\,\textrm{kpc}\approx6''$) aperture. In fact, the left panel of Fig.~\ref{fig:fgal} demonstrates that the ratio of these two equivalent widths is almost identical to the inferred $f_{\textrm{gal}}^\textrm{Ly$\alpha$}$ value from the MCMC fit for every sample.

This correspondence can be understood from the fact the $0\farcs6\approx5\,$kpc smaller aperture radius contains the majority of the ``galaxy'' emission in both the \lya\ line and the continuum, while 
the ``total'' \lya\ emission (i.e., the denominator in the calculation of $f_{\textrm{gal}}^\textrm{Ly$\alpha$}$) is defined in our model to be the emission within the same $R=50$ kpc radius used to define $\ewlya^{50}$. While the correspondence may therefore appear tautological, it provides a useful check that the integrated flux of the PSF-convolved exponential profiles yield very similar descriptions of the small-scale and large-scale \lya\ distribution to their direct measurements in circular apertures. Furthermore, the variation in this ratio illustrates the qualitative differences in the growth of \ewlya\ with radius among the high-\ewlya\ samples that have tightly clustered values of $f_{\textrm{gal}}^\textrm{Ly$\alpha$}\approx40\%$ vs. those with smaller and/or negative values of $f_{\textrm{gal}}^\textrm{Ly$\alpha$}$.

The relationship between the \ewlya\ values on these two different scales is shown in the right panel of Fig.~\ref{fig:fgal}. The two measurements are strongly correlated, but in the context of the left panel it is clear that their ratio and  $f_{\textrm{gal}}^\textrm{Ly$\alpha$}$ vary significantly across the samples. The samples with strong central \lya\ emission have very similar values of $f_{\textrm{gal}}^\textrm{Ly$\alpha$}$, such that $\ewlya^{50}\approx (2-3)\times \langle\ewlya\rangle$ could be inferred to be a general description of the variation of \ewlya\ with aperture. However, such a relationship would under-predict the observed large-scale \lya\ emission from the galaxy samples with central \lya\ absorption. That is, it would be difficult to infer the true relationship between the large- and small-scale \ewlya\ values from measurements of strong (i.e., centrally-dominated) \lya-emitters alone. These two plots together thus indicate the importance of including samples with central \lya\ absorption in comprehensive studies of \lya\ emission. 

\subsection{Central galaxy scale lengths}
\label{sec:centralcomparison}

\begin{figure*}
\centering
\includegraphics[height=0.32\textwidth]{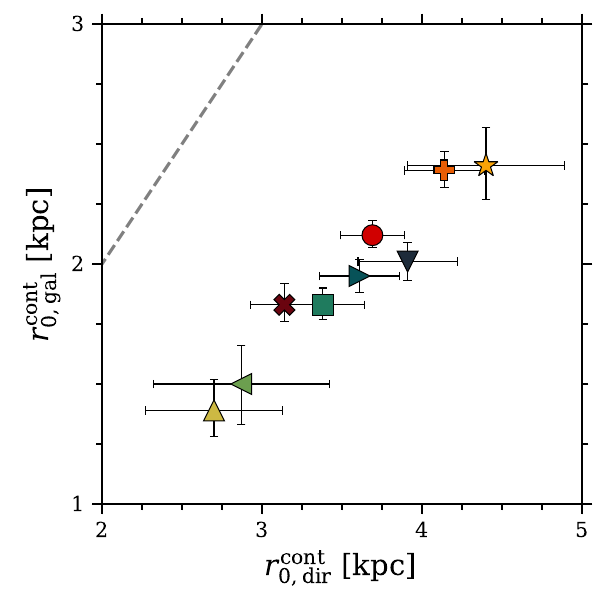}\hspace{2mm}
\includegraphics[height=0.32\textwidth]{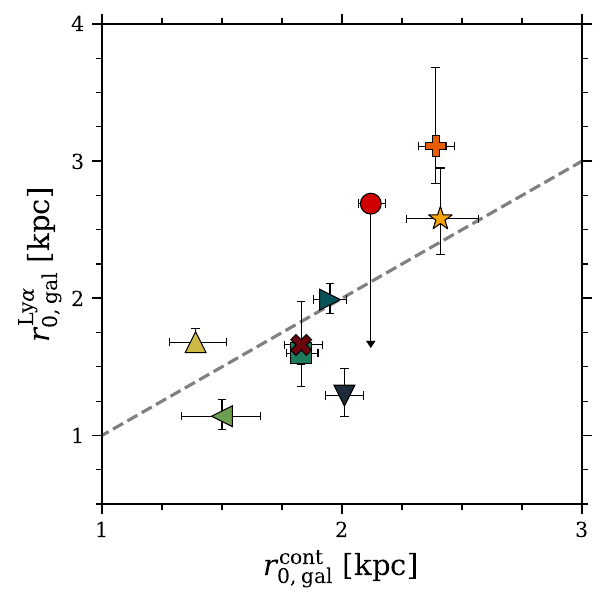}\hspace{2mm}
\includegraphics[height=0.32\textwidth]{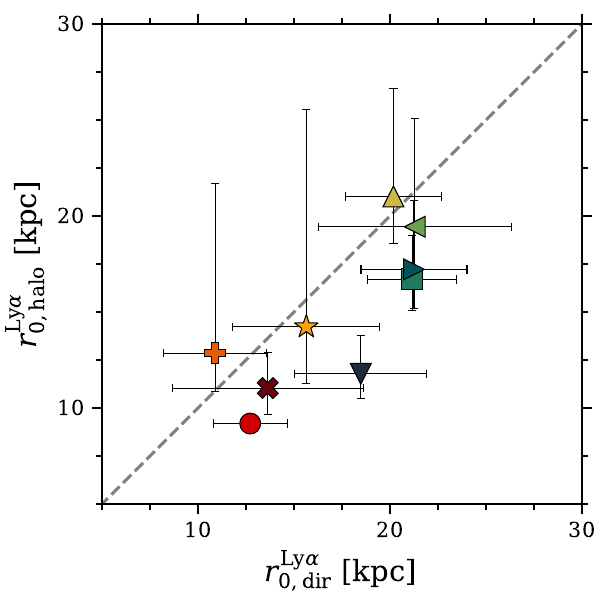}
\caption{Comparison of parameter values extracted from the direct exponential fits to the 1D emission profiles and those obtained via our forward-modeling framework as described in Secs.~\ref{sec:expfits}--\ref{sec:mcmcfits}. In each panel, the gray dashed line indicates the 1:1 relation for the plotted quantities. The KBSS All value in the center panel is given as a 2$\sigma$ (95\%) upper limit, and the error bars represent 1$\sigma$ uncertainties. Points are described by the legend in Fig.~\ref{fig:fgal}.}
\label{fig:parameter_comparison}
\end{figure*}

As displayed in Fig.~\ref{fig:parameter_comparison} and Table~\ref{table:fits}, the continuum scale lengths fit by the forward-modeling approach $r_{0,\textrm{gal}}^\textrm{cont}$ are generally smaller by $\sim$50\% than the values of $r_{0,\textrm{dir}}^\textrm{cont}$ measured from the direct fits to the 1D profiles %(Sec.~\ref{sec:expfits}).
(Fig.~\ref{fig:parameter_comparison}, left panel). This offset reflects the fact that the scale lengths inferred from the forward modeling describe the intrinsic light profile before being broadened by the PSF.

The inferred intrinsic central scale lengths for the \lya\ and continuum emission, $r_{0,\textrm{gal}}^\textrm{Ly$\alpha$}$ and $r_{0,\textrm{gal}}^\textrm{cont}$ are generally consistent with each other (Fig.~\ref{fig:parameter_comparison}, center panel), suggesting the central \lya\ emission component generally tracks the distribution of UV-luminous stars in the galaxy. Interestingly, this tight relationship between $r_{0,\textrm{gal}}^\textrm{Ly$\alpha$}$ and $r_{0,\textrm{gal}}^\textrm{cont}$ holds even for the KBSS Abs and KBSS All stacks, where the amplitude of the central \lya\ component is negative. This correspondence suggests that the dominant \lya\ absorption in these galaxies  occurs on similar spatial scales to the distribution of hot stars, at least to the extent that these scales can be probed by our seeing-limited imaging. To our knowledge, this result is the first time that a correspondence has been measured\footnote{This similarity in scale is visible in Fig.~6 of \citet{steidel2011}, although the \lya\ absorption and continuum emission profiles are not quantitatively compared in that paper.} between the spatial scales of stellar emission and \lya\ absorption in galaxies at high redshift, although this relationship may be expected to the extent that young stars track the distribution of high \ion{H}{1} column density gas on kpc scales. The correspondence of the physical scales of continuum emission and \lya\ absorption may also rely on the fact that \lya\ absorption is less easily detectable beyond the projected spatial distribution of the stars whose continuum emission is being absorbed. 

The KBSS All stack has a negative galaxy component with a small amplitude that is consistent with being a point-like component; the value of $r_{0,\textrm{gal}}^\textrm{Ly$\alpha$}$ is given as a 2$\sigma$ (95\% confidence) upper limit in Table~\ref{table:fits}. Because the center of the galaxy profile is the point with the greatest variation among the individual \lya\ images that are combined into this stack, and because the KBSS All sample includes the largest variation in central emission and absorption of any galaxy subsample, the shape of the inner profile for this stack is particularly sensitive to the sigma-clipping, scaling, and weighting of the individual images. As noted in Sec.~\ref{sec:stacks}, the KBSS All stack is the unweighted mean of the unscaled images in the KBSS All sample, and we employ a conservative 4$\sigma$ clipping algorithm. Lower clipping thresholds tend to censor emission from bright objects, which would tend to decrease the residual flux in the profile center and drive the amplitude of the best-fit galaxy component (and the related parameter $f_{\textrm{gal}}^\textrm{Ly$\alpha$}$) to more strongly negative values. The inferred scale length of the galaxy component varies less significantly with changing in stacking method than does its amplitude, but there are small changes owing to variation in the detailed shape of the transition from rising surface brightness to falling with increasing radius, which affect both the galaxy scale length $r_{0,\textrm{gal}}^\textrm{Ly$\alpha$}$ as well as $r_{0,\textrm{halo}}^\textrm{Ly$\alpha$}$. We use a fixed set of stacks for our analysis, but marginalizing over a range of stacking methods (i.e., by including various sigma-clipping thresholds and scaling parameters as nuisance parameters) would tend to increase the inferred variation in the scale lengths, particularly for the KBSS All and KBSS Abs fits as discussed further in Sec.~\ref{sec:halocomparison}.

\subsection{Halo scale lengths}
\label{sec:halocomparison}

The halo components of the \lya\ profiles generally have similar scale lengths to those inferred from the direct fits to the 1D surface brightness distributions (Fig.~\ref{fig:parameter_comparison}, right panel). There is a small systematic offset between the two techniques, with the forward-modeled scale-lengths being slightly smaller on average, but the scale lengths inferred for each sample are generally statistically consistent between the two measures.

The uncertainties on the forward-modeled scale lengths are generally larger, owing to the additional freedom in the constant background term---generally, the large-scale halo amplitude and slope are highly dependent on variations in the background, and those parameters exhibit strong covariance in the posterior distributions.

An exception to this trend is the KBSS All stack, and to a lesser extent the KBSS Abs stack, for which the uncertainties on $r^{\textrm{Ly}\alpha}_\textrm{0,halo}$ are smaller than for $r^{\textrm{Ly}\alpha}_\textrm{0,dir}$. These two stacks have the smallest fit values of $r^{\textrm{Ly}\alpha}_\textrm{0,halo}$, and they also are the two models with negative ``galaxy'' components (i.e., negative values of $f^{\textrm{Ly}\alpha}_\textrm{gal}$). The steepness of the halo profiles in these stacks makes them less degenerate with changes in the constant background; likewise the negative central absorption causes variations in the halo and galaxy scale lengths to be less degenerate with one another than they are among models with positive amplitudes for both exponential components. As noted in Sec.~\ref{sec:centralcomparison}, including a range of stacking methods as a source of uncertainty in our fit would tend to increase the uncertainty in our parameter estimates, so the uncertainties presented here should be interpreted in the context of a fixed choice of stacking methodology.

\section{Discussion}
\label{sec:discussion}

\subsection{Comparison to other KBSS studies}
\label{sec:kbsscomparison}

Here we compare to previous studies of \lya\ halos among galaxies in the KBSS, including the earlier analysis by \citet{steidel2011} of similar galaxy stacks from Keck/LRIS and Palomar/LFC, as well as more recent works by \citet{chen2021} and \citet{erb2023} that present KCWI measurements of \lya\ emission around KBSS galaxies.

\subsubsection{KBSS sample comparisons}
\label{sec:steidelsample}
As noted in Sec.~\ref{sec:kbss-sample}, the KBSS galaxies included in our sample overlap with those considered by \citet{steidel2011}, but there are also significant differences in our observations and methodology. Specifically, the sample of \citet{steidel2011} includes 27 galaxies in the HS1549+1919 field that are also included in our own sample of 29 KBSS galaxies in that field; we also use the same deep narrowband image from Keck/LRIS described in that earlier work. However, our sample omits the 65 galaxies in the SSA22a and HS1700+6416 fields studied by \citet{steidel2011}. While we do include several KBSS-\lya\ and KBSS galaxies from the HS1700-6416 field in this analysis, our \lya\ selection and imaging uses a new Keck/LRIS NB4535 image ($\lambda_\textrm{cen}\approx4535$\AA) that is sensitive to \lya\ emission at $z\approx2.75$. Conversely, \citet{steidel2011} selected galaxies in a lower-redshift overdensity at $z\approx2.3$ in this same field, which were observed using a $\lambda_\textrm{cen}\approx4018$\AA\ \lya\ narrowband image from Palomar/LFC; the two sets of galaxies in this field are thus entirely distinct. The SSA22a narrowband image described by \citet{steidel2011} is constructed from a combination of Keck/LRIS and Subaru/SuprimeCam images, selecting \lya\ emission at $z=3.1$.  In our work, we have elected to focus on nine fields that span a narrow range of redshifts ($2.55-2.83$; Table~\ref{table:fields}) and we use a single instrument (Keck/LRIS-B) for all of the \lya\ and continuum imaging in order to optimize the consistency of our observations while also maintaining the statistical power to average over field-to-field variations.

\citet{chen2021} present Keck/KCWI observations of 59 KBSS galaxies across 11 fields. The galaxies were selected to maximize the number of objects that could be observed in a limited number of pointings while also spanning the the full population of KBSS galaxies in their stellar masses ($10^9\lesssim M_*/M_\odot\lesssim 10^{11}$), star formation rates ($1\lesssim\log(\mathrm{SFR}/[M_\odot\,\mathrm{yr}^{-1}])\lesssim 100$), \lya\ emission strength ($-30\lesssim \mathrm{EW}_{\textrm{Ly}\alpha}/\textrm{\AA}\lesssim 70$), and rest-optical properties.

\citet{erb2023} present similar Keck/KCWI observations for a sample of 12 relatively low-mass ($M_*\sim10^{9}\,M_\odot$) galaxies across eight KBSS fields, with each galaxy previously known to exhibit extreme nebular emission line ratios and high \lya\ equivalent width indicative of strong ionization. These low masses and high excitations are broadly similar to those of the KBSS-\lya\ sample of \citetalias{trainor2016}, although generally with higher star formation rates in  the \citet{erb2023} sample (median SFR = 14 M$_\odot$/yr$^{-1}$) than in the KBSS-\lya\ (median SFR = 5 M$_\odot$/yr$^{-1}$). The similarities between these \lya-selected and nebular-emission-line galaxy samples are discussed further in both \citetalias{trainor2016} and \citet{erb2016}.

\subsubsection{Comparison of KBSS halo measurements}
\label{sec:steidelresults}

As described in Sec.~\ref{sec:intro}, \citet{steidel2011} measured the exponential scale lengths of their \lya\ and continuum surface brightness profiles via a direct fit of an exponential function, omitting the inner portion of each profile where the distribution diverges visually from a pure exponential. They use this technique to fit the their total stack of 92 galaxies, as well as stacked subsets selected on the basis of net \lya\ emission, absorption, or LAE-like emission (EW$_{\textrm{Ly}\alpha}>20$\AA) analogous to the KBSS stacks considered here. For these stacks, they derive \lya\ halo scale lengths of 25.2 kpc (``All''), 25.6 kpc (``Em''), 20.8 kpc (``Abs''), and 28.4 (``LAE''). Their fits to the stellar continuum stacks for these same galaxy subsamples inferred scale lengths of 3.4 kpc, 2.9 kpc, 4.5 kpc, and 2.9 kpc, respectively.

Our direct exponential fits (Sec.~\ref{sec:expfits}) are designed to mimic the methods employed by \citet{steidel2011}, and we generally find smaller \lya\ halo scale lengths and smaller or similar continuum scale lengths (Figs.~\ref{fig:1Dcont_kbss}~\&~\ref{fig:1Dlya_kbss}; Table~\ref{table:fits}) for each of our analogous samples. In particular, the typical peak surface brightness of our \lya\ halo profiles at the point where they break from the central ``galaxy'' component are similar to those reported by \citet{steidel2011} ($S_{\textrm{Ly}\alpha}\approx 10^{18}$ erg s$^{-1}$ cm$^{-2}$ arcsec$^{-2}$), but our \lya\ halo scale lengths are 50$-$70\% of the values reported in that work.

One likely source of the differences in these results is simply sample variance among the selected fields: while \citet{steidel2011} found statistically consistent surface brightness distributions among their three fields, a visual inspection of the profile from the HS1549 field suggests that it may also be consistent with a slightly steeper profile than the other two included fields. A full analysis of field-to-field variations in the \lya\ halo properties is reserved for future work. In addition, the inner cutoff of our exponential fits at 20 kpc is smaller than the radius at which the \citet{steidel2011} profiles diverge from the exponential fits, such that applying our range of fitted radii to the \citet{steidel2011} stacks would likely result in smaller scale lengths than those reported in that work. Finally, our sky-subtraction and stacking procedures differ between the two works, with our current analysis using a sigma-clipping algorithm in the creation of our stacks, whereas the earlier work employed a straight mean after masking background continuum sources in the individual images. \citet{steidel2011} note that replacing their mean-combined stacks with median-combined stacks resulted in a smaller fit scale length ($r_0=17.5$ kpc), so the more aggressive outlier-rejection we have applied in this analysis may be another contributor to our smaller inferred values.

Comparing the two sets of measurements as a whole, we note here that our direct fits to the large-scale \lya\ surface brightness profiles are broadly consistent with the results of our forward-modeling techniques Our measurements here reflect the increased focus on background subtraction and large-scale PSF effects that have been noted by other studies over the last decade, but it does not appear that the \citet{steidel2011} results were strongly biased by their instrumental PSF.  Planned future analyses of the environmental dependence and field-to-field variations among our \lya\ halo profiles may shed more light on whether the differences in inferred scales are due to physical effects or to the remaining systematic differences in methodology.

\begin{figure}
\includegraphics[width=0.47\textwidth]{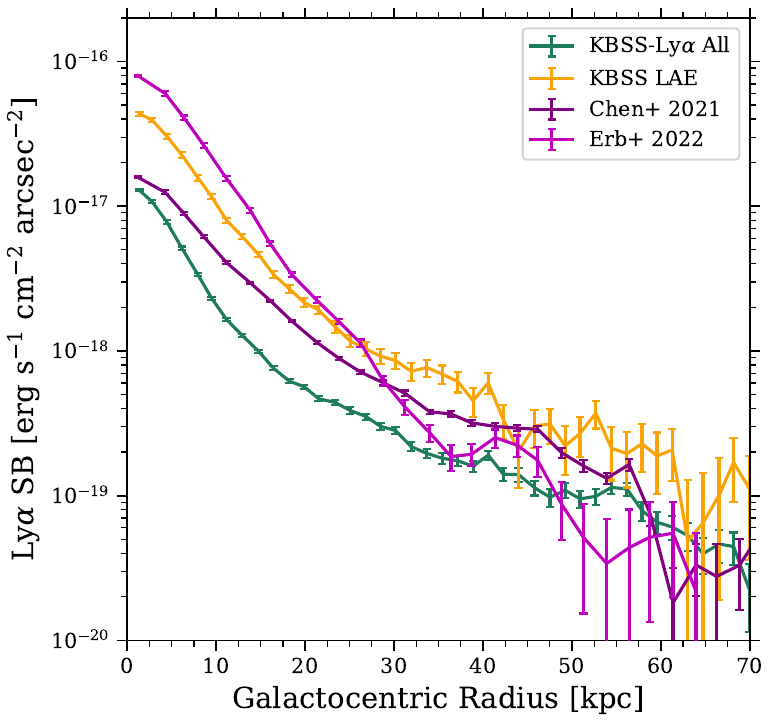}
\caption{Azimuthally-averaged \lya\ surface brightness profiles for galaxy samples selected in KBSS fields, including KCWI observations by \citet{chen2021} and \citet{erb2023} and the KBSS-\lya\ All and KBSS LAE samples. All of the samples have generally similar \lya\ halo profile shapes, although with different contributions from the inner ``galaxy'' and outer ``halo'' components.} \label{fig:lris_kcwi_lya_empirical}
\end{figure}

As a further comparison to recent measurements of \lya\ profiles of galaxies in the KBSS fields, Fig.~\ref{fig:lris_kcwi_lya_empirical} displays surface brightness profiles of stacked KCWI observations from \citet{chen2021} and \citet{erb2023}. The \citet{chen2021} stack is described in that paper, while we construct the \citet{erb2023} stack using individual \lya\ images extracted from the KCWI datacubes, which are then averaged in the same manner as the LRIS stacks described in Sec.~\ref{sec:stacks}. The KBSS-\lya\ ``All'' and KBSS ``LAE'' profiles are shown for comparison, as their centrally-dominated \lya\ emission is most similar to that of the two KCWI stacks. Despite the differences in \lya\ luminosities, selection criteria, and the instruments used to observe these samples, all of the the observed profiles are generally consistent at large radii. The \citet{erb2023} sample has a particularly high fraction of its total \lya\ emission coming from the inner ``galaxy'' component rather than the outer halo, such that the transition from the inner to outer profiles is less clear. Given that the \citet{erb2023} sample is selected to have strong nebular line emission in the central galaxy, it is perhaps not surprising that the \lya\ emission from these galaxies is centrally dominated as well.

\subsection{Comparison to individual and stacked measurements from VLT/MUSE}
\label{sec:musecomparison}

Recent measurements with VLT/MUSE have characterized \lya\ halo profiles around individual galaxies using integral-field spectroscopy: \citet{wisotzki2016} presented 26 galaxies of which 21 have detectable halos suitable for profile fitting, and \citet{leclercq2017} reported a significantly larger sample of 145 galaxies, again with $\sim$80\% having sufficient halo detections for profile fitting. Being detected directly by their \lya\ emission in deep VLT/MUSE observations, the galaxies in both of these samples are qualitatively similar to the continuum-faint KBSS-\lya\ sample of LAEs, although the MUSE samples reach lower luminosities and \lya\ surface brightnesses and are limited to $z>3$ due to the short-wavelength cutoff of VLT/MUSE at $\lambda_\textrm{obs}\sim5000$\AA.

\begin{figure*}
\centerline{
\includegraphics[height=0.4\textwidth]{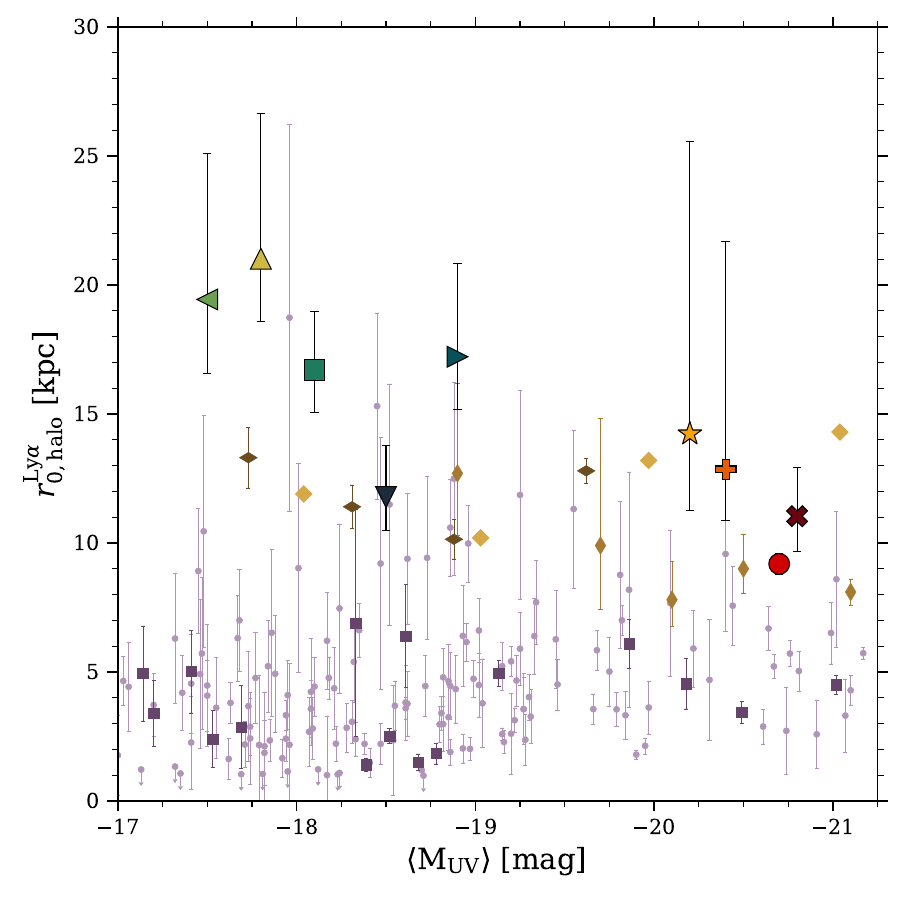}
\includegraphics[height=0.4\textwidth]{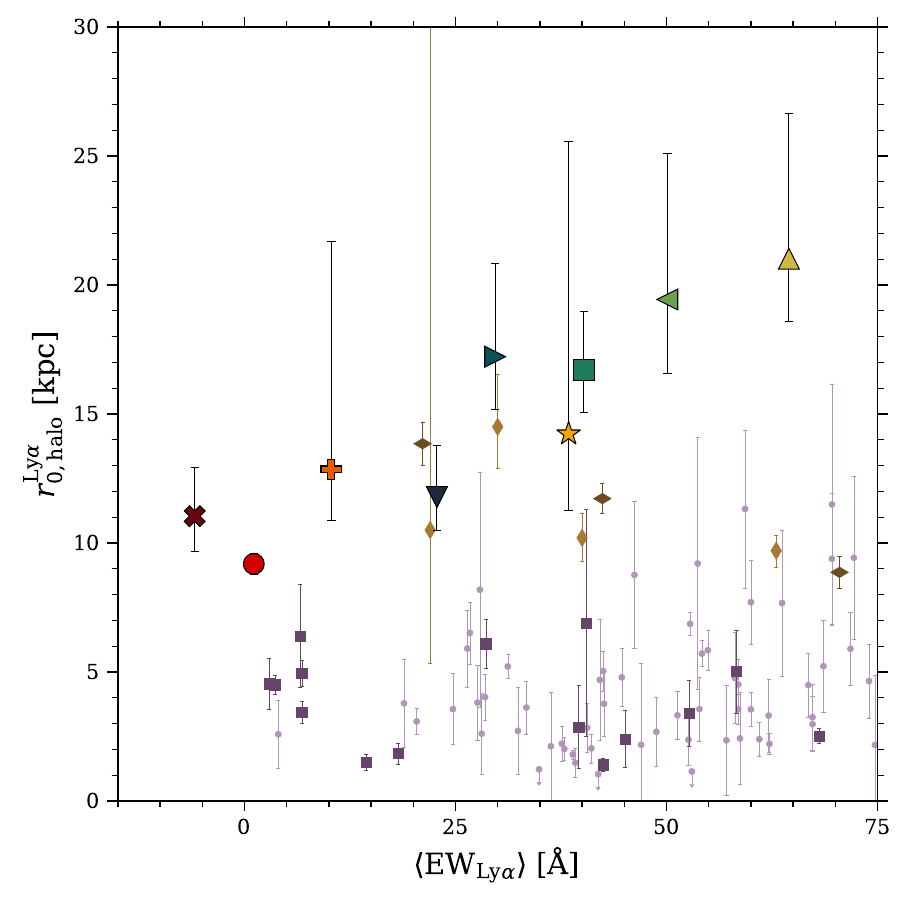}
\includegraphics[height=0.4\textwidth]{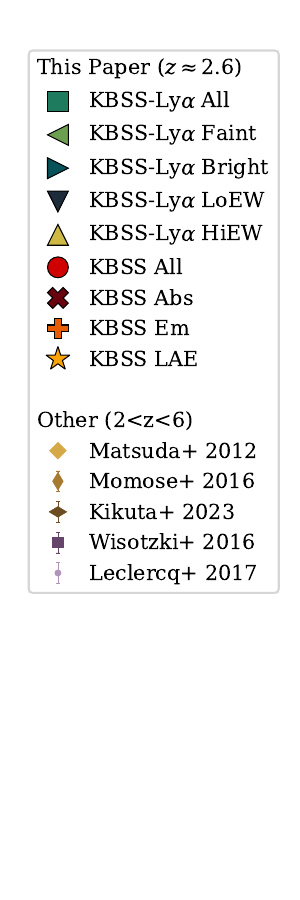}
}
\caption{Best-fit \lya\ halo scale lengths obtained via our forward-modeling technique (Sec.~\ref{sec:mcmcfits}) as a function of the median UV luminosity M$_\textrm{UV}$ (left panel) and median \lya\ equivalent width (right panel) for the stacked galaxy samples presented in this paper as well as several measurements from the literature for stacks or individual measurements.  The scale lengths display weak trends among our samples, such that larger scale lengths are associated with higher \ewlya\ and lower M$_\textrm{UV}$. Comparison samples are given in the legend, including stacked measurements from 
\citet{matsuda2012} at $z=3.1$, \citet{momose2014,momose2016} at $2<z<6$, and \citet{kikuta2023} at $z=2.8$. We also include individual VLT/MUSE halo measurements at $3<z<6$ from \citet{wisotzki2016} and \citet{leclercq2017}.}\label{fig:r_vs_MUV_EWLya}
\end{figure*}

\begin{figure*}
\centerline{
\includegraphics[height=0.4\textwidth]{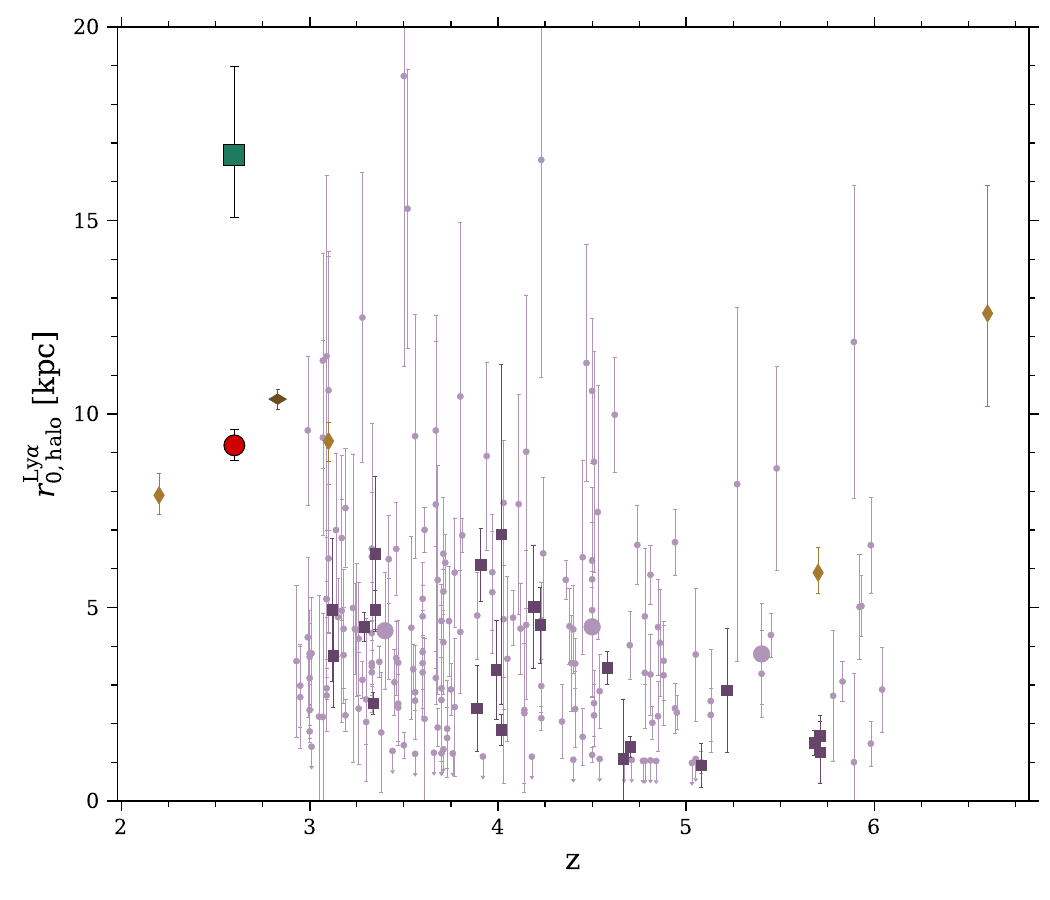}\hspace{5mm}
\includegraphics[height=0.4\textwidth]{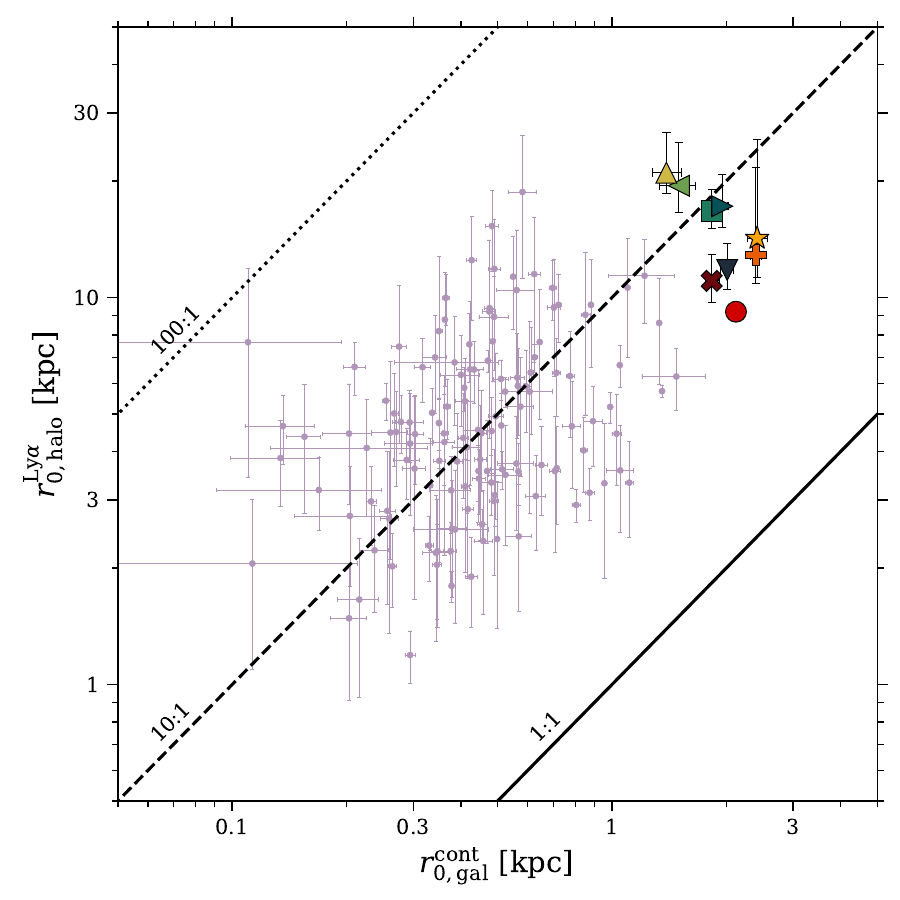}
}
\caption{Best-fit \lya\ halo scale lengths as a function of redshift  (left panel) and continuum scale length (right panel) for the KBSS and KBSS-\lya\ samples as well as comparison samples as in Fig.~\ref{fig:r_vs_MUV_EWLya}. The oversized lilac circles represent redshift-binned stacked from \citet{leclercq2017}. Because all of the galaxy subsamples in this paper have the same average redshift, only the KBSS All and KBSS-\lya\ All stack measurements are shown in the left panel.}\label{fig:r_vs_z_rcont}
\end{figure*}

Both sets of VLT/MUSE measurements are included in the distributions of fitted halo properties and their trends with M$_\textrm{UV}$ and \ewlya\ in Fig.~\ref{fig:r_vs_MUV_EWLya}.  The VLT/MUSE halo profiles are fit using a similar double-exponential, PSF-convolved model as is used in the forward-modeling method of this paper (Sec.~\ref{sec:mcmcfits}). When plotting the \ewlya\ values for the MUSE/VLT samples in the figures here, we use the central \ewlya\ values as reported in those papers in order to compare to the median individual \ewlya\ measurements for each of our subsamples; as discussed above, the \ewlya\ values for both our samples and the VLT/MUSE samples are greater when measured using larger apertures. Weak trends are also visible in both cases, in the sense that galaxy samples having fainter UV continua and/or larger central EW$_\textrm{Ly$\alpha$}$ tend to have slightly larger best-fit halo scale lengths. 

The most significant difference among the VLT/MUSE measurements and those presented in this paper are the significantly smaller halo scale lengths measured by VLT/MUSE ($r_\textrm{0,halo}\approx5\,$kpc).  A key result of \citet{leclercq2017} is the general lack of any apparent trend between the halo scale length and other galaxy properties, in contrast to the results we find here. For this reason, the differences in typical scale lengths between the \citet{leclercq2017} sample and the stacks here are most prominent for the UV-faint (M$_\textrm{UV}\lesssim -19$) and high-EW$_{\textrm{Ly}\alpha}$ (EW$_{\textrm{Ly}\alpha}\gtrsim 20$\AA) galaxies where we find the largest halos.

Fig.~\ref{fig:r_vs_z_rcont} displays the relationship between halo scale length and redshift for the samples included in Fig.~\ref{fig:r_vs_MUV_EWLya}. In both cases, no trend is visible among our samples alone, but our samples are consistent with the negative trend in $r^{\textrm{Ly}\alpha}_\textrm{0,halo}$ vs. $z$ among the \citet{wisotzki2016} measurements and the positive trend in $r^{\textrm{Ly}\alpha}_\textrm{0,halo}$ vs. $r^{\textrm{cont}}_\textrm{0,gal}$ from \citet{leclercq2017}. However, the MUSE and KBSS studies are difficult to compare directly owing to the lack of any points in the VLT/MUSE samples with comparable redshifts or sizes to those in our sample. It is possible this trend is related to the strong $(1+z)^4$ surface brightness dimming that makes it challenging to detect faint, extended halo components at higher redshifts. However, the deep MUSE spectroscopy seems to clearly distinguish a central galaxy and outer halo component in many cases, and other studies discussed below find trends of increasing halo scale with redshift, so any systematic effect of surface-brightness dimming on the measured halo profiles is unclear.

Notably, the selection on \lya\ emission used by \citet{wisotzki2016} and \citet{leclercq2017} prevents the inclusion of objects with central \lya\ absorption, so we are unable to compare the KBSS-All and KBSS-Abs samples to VLT/MUSE measurements. However, the more recent VLT/MUSE halo measurements by \citet{kusakabe2022} include UV-selected galaxies with a range of central \ewlya\ values. While those authors do not explicitly fit an exponential profile to the halos, their mean radial surface brightness profile is visually consistent with the mean profile from \citet{wisotzki2016}, again indicating a lack of dependence on \ewlya\ and suggesting that halo scale-lengths measured from the \citet{kusakabe2022} sample would likewise be smaller than those inferred by our analysis.

\subsection{Comparison to stacked measurements from Subaru/Suprime-Cam}
\label{sec:subarucomparison}

Fig.~\ref{fig:r_vs_MUV_EWLya} also includes previous studies conducted with Subaru/Suprime-Cam \citep{matsuda2012,momose2014,momose2016} and Subaru/Hyper Suprime-Cam (HSC; \citealt{kikuta2023})  where appropriate. These studies were based on stacked narrow-band \lya\ imaging of \lya-selected galaxies at $2\lesssim z\lesssim 6$, thus overlapping more closely with the observational technique and redshift distribution of our study. In particular,  the observations of \citet{kikuta2023} are entirely localized to HS1549+1919 field at $z=2.83$. This is the same field included in both this study and in \citet{steidel2011}, although the Subaru/HSC imaging presented by \citet{kikuta2023} covers a much wider field of view and is also substantially deeper, allowing them to select 3490 \lya-emitting galaxies in that field alone.  However, all the Subaru samples also lack galaxies with central \lya\ absorption, and the halo scale-length measurements from \citet{matsuda2012} and \citet{momose2014,momose2016} are made using a direct fit to the halo profile similarly to \citet{steidel2011} and Sec.~\ref{sec:expfits} of this paper, rather than using a forward-modeling approach that directly accounts for PSF effects. \citet{kikuta2023} fit halo profiles with a double-exponential model\footnote{\citet{kikuta2023} also fit their halo profiles with an alternative power-law model (again convolved with the instrumental PSF) motivated by analytical predictions of \citet{kakiichi2018}. Because none of the other comparison studies or our on work include this approach, we include only the double-exponential fit results from \citet{kikuta2023} here.} convolved with the instrumental PSF, similar to our forward-modeling approach here. 

The scale lengths derived by the Subaru studies are $r_{0}\approx8-15\,$kpc, generally falling between those measured in our study and in the VLT/MUSE studies. However, unlike the VLT/MUSE results summarized above, the Subaru results exhibit significant variation in halo scale among galaxy subsamples binned by various galaxy or environmental properties. Most significantly, both \citet{matsuda2012} and \citet{kikuta2023} found a correlation between galaxy overdensity and halo scale length, in the sense that galaxies in high-density regions on Mpc scales are associated with larger halos. \citet{momose2016} also found weak halo trends with M$_\textrm{UV}$ and central \ewlya, as well as a stronger correlation with central \lya\ luminosity, with UV-bright, \lya-faint, and low-\ewlya\ galaxies having the largest halos (up to $\sim$15 kpc). Thus while the range of halo scale lengths in the Subaru/Suprime-Cam studies is comparable to the range measured here, we find opposing trends for halo size vs. M$_\textrm{UV}$ and vs. central \ewlya: we find the largest halos among UV-faint and high-\ewlya\ galaxies, where \citet{momose2016} found the smallest halos. In contrast, \citet{matsuda2012} and \citet{kikuta2023} found no significant relationship between halo scale length and M$_\textrm{UV}$ at all.

\section{Summary \& Conclusions}
\label{sec:summary}

In this work, we have presented the imaging data and photometry for the KBSS-\lya, including the first measurements of extended \lya\ halos in stacks of 734 \lya-selected galaxies and new measurements of \lya\ halos in stacks of 119 more luminous continuum-selected galaxies in the KBSS fields at $z\sim2.6$. We split each survey sample into multiple subsets based on central \lya\ equivalent width and UV continuum luminosities in order to probe the dependence of \lya\ halo morphology on these properties. Using multiple techniques to characterize the radial surface brightness distributions of \lya\ and UV continuum emission, we find the following:
\begin{enumerate}
\item All of the galaxy image stacks exhibit extended \lya\ emission at radii well beyond that of their continuum light profiles, extending to scales comparable to the virial radius of the host dark-matter halos and the circumgalactic media of these galaxies. (Sec.~\ref{sec:halos}, Figs.~\ref{fig:2DPanel_kbsslya}$-$\ref{fig:1Dcont_kbss}). 
\item Regardless of the strength of their central \lya\ or continuum emission properties, all of our galaxy stacks exhibit net positive \lya\ emission within apertures of $R\approx 50\,$kpc, including galaxies selected as central \lya\ absorbers ($\theta\approx 6''$; Sec.~\ref{sec:curveofgrowth} and Fig.~\ref{fig:EWcog}).
\item Using direct exponential fits to the 1D radial \lya\ and continuum emission profiles, we find that the stacks of continuum-faint, high-\ewlya\ KBSS-\lya\ galaxies exhibit extended \lya\ emission with typical halo scale lengths $r^{\textrm{Ly}\alpha}_{0,\textrm{dir}}\approx 20\,$kpc. Using the same method, we find slightly smaller halo scale lengths for the continuum-bright KBSS galaxy samples, with $r^{\textrm{Ly}\alpha}_{0,\textrm{dir}}\approx 10-15\,$kpc. (Sec.~\ref{sec:expfits}, Figs.~\ref{fig:1Dlya_kbsslya}$-$\ref{fig:1Dcont_kbss}).
\item Using a second technique in which we fit a dual exponential profile with inner ``galaxy'' and outer ``halo'' components convolved with the empirical PSF, we generally find consistent or slightly smaller halo scale lengths on average, with  $r^{\textrm{Ly}\alpha}_{0,\textrm{halo}}\approx 10-20\,$kpc for both the KBSS and KBSS-\lya\ samples. (Sec.~\ref{sec:mcmcfits}, Fig.~\ref{fig:mcmc_plot_grid}).
\item Our forward-modeling technique is able to fit image stacks exhibiting central \lya\ emission or central \lya\ absorption using the same dual-exponential model. The exponential scale length of the central ``galaxy'' \lya\ component is generally consistent with the exponential scale length of the continuum profile for a given galaxy sample, regardless of whether the amplitude of the central \lya\ component is positive or negative.
\item Among our subsamples, we find a weakly negative relationship between the scale length of the \lya\ halo profiles and the UV continuum luminosity, as well as a weak positive correlation with the central \lya\ equivalent width (\ewlya), in the sense that our continuum-faint, high-\ewlya\ subsamples have the largest \lya\ halo scale lengths. These results are in contrast to recent results from VLT/MUSE and Subaru/Suprime-Cam, which found smaller halos and no (or opposite) trends with UV-continuum luminosity and \ewlya.
\end{enumerate}

Given our results in the context of the existing literature, it remains difficult to discern any clear trends between halo scale length and other physical galaxy properties. While these various works have substantial overlap in their sampled distributions of galaxy redshifts, continuum luminosities, and \ewlya, as well as shared observational techniques and similar analysis, the primary factor predicting large or small scale-length measurements seems to be the telescope and instrument used to perform the observations. It may be necessary to resolve the remaining differences in methodology (e.g., the details of stacking strategies, sky-subtraction, and incorporation of PSF effects) in order to compare these measurements with confidence.

One remaining physical driver of the variation in scale length is its possible relationship with a galaxy's local Mpc-scale overdensity discussed above. \citet{momose2016} argue that the trend of increasing \lya\ halo scales with higher galaxy density presented in \citet{matsuda2012} is consistent with their own results and those of \citet{steidel2011}, given the varying overdensities among the fields sampled by those studies. Likewise, \citet{kikuta2023} find a significant increase in halo scale length at the highest overdensities, from rising from $r^{\lya}_\textrm{halo}\sim10\,$kpc at $\delta\lesssim2$ to $r^{\lya}_\textrm{halo}\sim45\,$kpc at $\delta\approx4$. While we do not directly test for a dependence on galaxy overdensity in this paper, we note that our narrowband \lya\ emissions are tuned to the redshifts of known bright quasars, several of which (including the HS1549 field also studied by \citealt{kikuta2023}) are known to occupy overdense regions on average---see further discussion in \citet{trainor2012,trainor2013} and \citet{kikuta2019}. Likewise, extended \lya\ nebulosities (i.e.,  \lya\ Blobs [LABs] or Enormous \lya\ Nebulae [ELANe]) have also been found to preferentially occupy overdense regions (e.g., \citealt{steidel2000,matsuda2004,prescott2008,cai2017}) and/or the vicinity of UV-luminous quasars (e.g., \citealt{cantalupo2014,hennawi2015,martin2014}), and \citet{steidel2011} argue that LABs may simply represent the highest-luminosity \lya\ halos similar to those found among typical galaxies. Thus the relatively large scale lengths measured within our study and in \citet{steidel2011} could conceivably be related to the enhanced matter densities and ionizing radiation fields of these atypical environments. However, we note that the KCWI measurements by \citet{chen2021} and \citet{erb2023} have similar halo profiles to our KBSS measurements despite not targeting overdense or quasar-associated environments; in addition, the luminosities and overdensities of the quasar fields included in this study vary widely. Future work will consider these environmental factors among the KBSS halo measurements.

\begin{acknowledgments}
 We are deeply indebted to those of Hawaiian ancestry on whose sacred mountain we are privileged to be guests. We also extend thanks to the staff of the W.M. Keck Observatory who keep the instruments and telescopes running effectively. R.F.T.\ 
 acknowledges support from the Pittsburgh Foundation (UN2021-121482) and the Research Corporation for Scientific Advancement (28289). D.K.E.\ is supported by the US National Science Foundation (NSF) through Astronomy \& Astrophysics grant AST-1909198. Undergraduate student co-authors of this work were supported by the Franklin \& Marshall Hackman program for undergraduate summer research, as well as by the Pittsburgh Foundation grant above.
\end{acknowledgments}

\software{Numpy \citep{numpy_harris2020},
Astropy \citep{astropy2013,astropy2018,astropy2022},
Matplotlib \citep{matplotlib2007},
Scipy \citep{scipy2020}
}

\bibliographystyle{aasjournalv7}
\bibliography{MyRefs}

\appendix

\section{Construction of the large-scale PSF}
\label{sec:psf}

The point spread function (PSF) used in our forward-modeling technique (Sec.~\ref{sec:mcmcfits}) is empirically determined from the stars in each field. Stars with a range of apparent magnitudes in the 18 BB or NB field images are averaged in order to characterize the PSF over the full range of angular scales relevant to our measurements of the \lya\ halos.
Bright stars are identified using the USNO B1.0 catalog \citep{monet2003} and visual identification. Additional stars with $19\lesssim B \lesssim 21$ are drawn from a catalog used for Keck/LRIS mask alignments; these fainter stars do not approach saturation even in their central pixels. In each field, a total of 17$-$46 stars are selected for PSF construction.

Each star is centroided using a 2D Moffat fit to the surface-brightness distribution after masking saturated pixels at the centers of the bright stars. Fits are performed using the Levenberg-Marquardt algorithm and Moffat functions implemented in the \texttt{astropy.modeling} library with the following functional form for the surface brightness distribution:
\begin{equation}
\textrm{S}(\theta)=A\left(1+\frac{\theta^2}{\gamma^2}\right)^{-\alpha}\label{eq:moffat}
\end{equation}
\begin{equation}
\theta=\left((x-x_0)^2+(y-y_0)^2\right)^{1/2}
\end{equation}

\noindent where $x_0$ and $y_0$ are the coordinates of the center of the star in the plane of the sky, $\gamma$ is the scale radius of the PSF in angular units, $\alpha$ is the power-law slope at large radii, and $A$ defines the normalization of the profile.

The shape and normalization of each profile are key to estimating its total flux, and these parameters are poorly constrained for individual stars. For this reason, we perform the Moffat fits iteratively. First, we fit the model allowing all parameters to vary freely in order to find the central coordinates $(x_0,\,y_0)$ and obtain an initial estimate of the $\gamma$ and $\alpha$ values for each star. 
A second fit is then performed 
after fixing $\gamma$ and/or $\alpha$ to the median values among the subset of stars in the same field image for which that parameter is well-constrained. In practice, the $\gamma$ values are estimated from the fainter stars with unsaturated centers, while $\alpha$ is more reliably measured from the brighter stars that have detectable flux in the outer PSF, particularly once the value of $\gamma$ is constrained to match that of the unsaturated stars. Stars in a given field with profiles inconsistent with the typical $\gamma$ and $\alpha$ values for other stars in the same field are omitted from the sample used to construct the PSF. 

To construct a single empirical PSF for a given NB or BB field image, the best-fit coordinates of each star are used to extract a $200\times 200$ pixel stellar image ($21''\times21''$) from the oversampled field images using the same algorithm used to extract the galaxy images (Sec.~\ref{sec:stamps}). Each stellar image is then normalized by its individual inferred total flux estimated from its corresponding constrained Moffat fit and then sigma-clipped with respect to the other stellar images from the same field image using \texttt{astropy.stats.sigma\_clip} after masking background sources and saturated pixels. The masked star images for each field image are then averaged with \texttt{np.ma.average}, weighting by the same inferred fluxes used for normalization. This process is repeated for all 18 field images, where the NB and BB images of the same field use the same set of stars to construct the mean PSFs for that field. At the largest radii, the empirical PSFs for several of the fields---particularly those with only a few bright stars---do not smoothly approach zero power as the PSF becomes dominated by  local errors in the sky subtraction or the incomplete masking of nearby objects. To correct this effect, we fit a Moffat profile to the average outer PSF ($5''\lesssim \theta \lesssim 10''$) for each field image and then smoothly transition from the empirical PSF to the fit model at large radii, such that the PSF at $\theta<5''$ is purely empirical and the PSF at $\theta>10''$ is a pure Moffat model.

\begin{figure*}
    \centering
    \includegraphics[width=.75\textwidth]{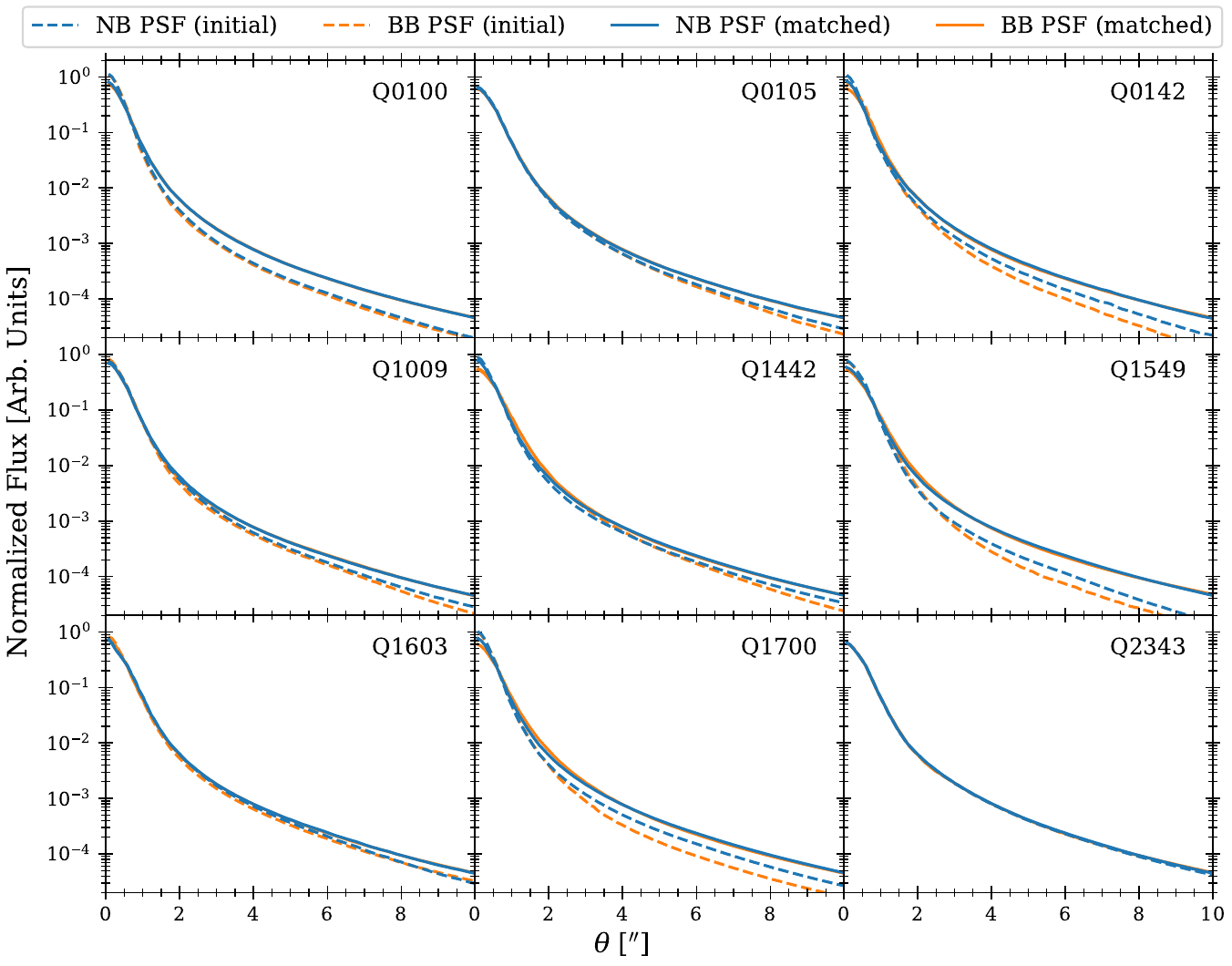}
    \caption{NB (blue) and BB (orange) PSFs plotted vs. projected angle for each field before (dashed) and after (solid) our iterative PSF-matching procedure. The matched PSFs show very close agreement across bands and fields.}
    \label{fig:PSFPanel}
\end{figure*}

The NB and BB PSFs constructed above have significant variation among the nine fields described by this work, with the Q2343 field in particular being observed under poorer seeing conditions (Fig.~\ref{fig:PSFPanel}). For this reason, all the PSFs are iteratively smoothed to match a ``broadest PSF" (BPSF) constructed from the full set of PSFs. We define the BPSF by comparing the cumulative radial flux distributions $F(<\theta)$ for each PSF and selecting the minimum value of $F(<\theta)$ at each $\theta$. The radial flux distribution of the BPSF corresponding to this cumulative distribution thus encircles no larger a fraction of its total flux at a given angle $\theta$ than any of the field-specific PSFs. In practice, the BPSF closely resembles the PSF for the Q2343 field. The smoothing kernel for each field image that causes it to most closely match the BPSF is then applied to the full corresponding field image.

\begin{figure*}
    \centering
    \includegraphics[width=0.75\textwidth]{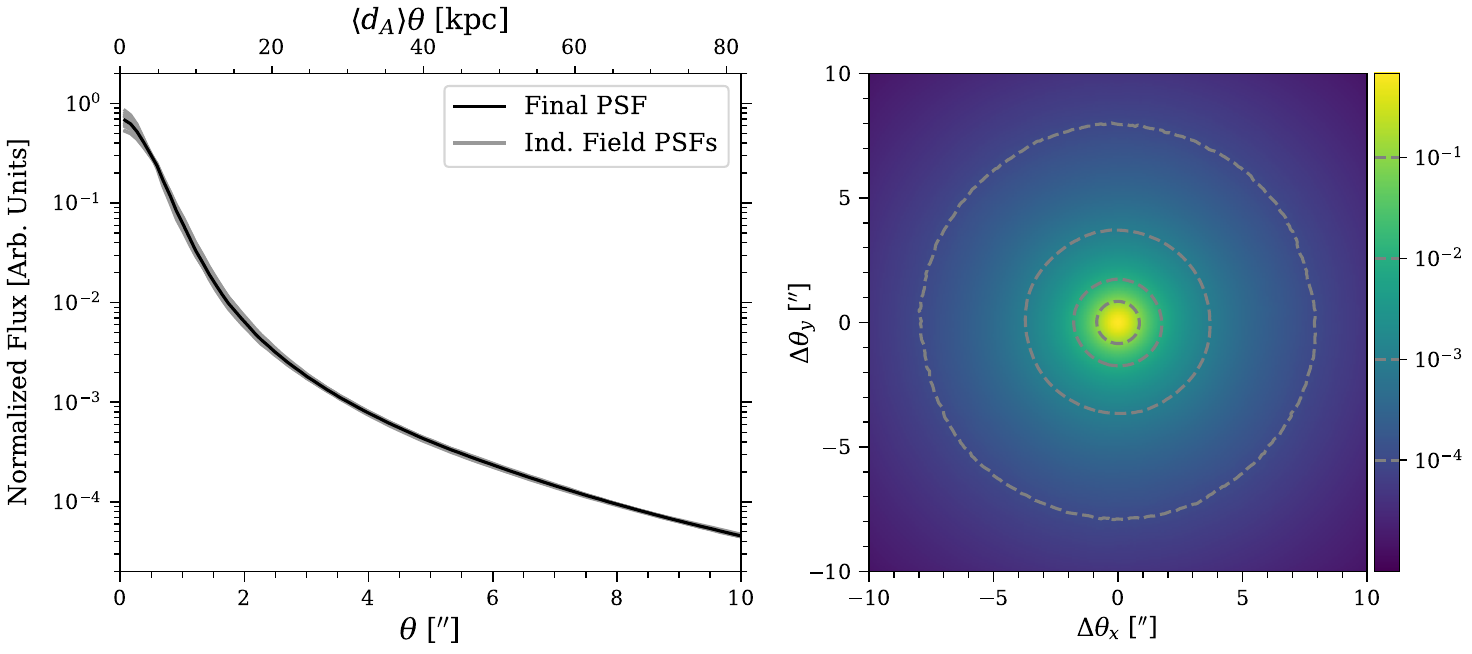}
    \caption{The final PSF used as the input to our forward-modeling procedure described in Sec.~\ref{sec:mcmcfits}. \textit{Left:} The final 1D PSF (black) plotted with the 18 individual PSFs for the PSF-matched field images (gray). The small width of the gray shading indicates the consistency of the matched PSFs. The top axis $\langle d_A\rangle \theta$ is the product of the projected angle with the angular diameter distance at the average field redshift, which is provided to facilitate comparison with the halo profiles presented in this paper. \textit{Right:} The final 2D PSF shown with logarithmic scaling to highlight its azimuthal symmetry. Gray contours trace isophotes relative to the peak (central) surface brightness as shown on the colorbar.}
    \label{fig:finalPSF}
\end{figure*}

The method described above ensures that the NB and BB images of a given field are PSF-matched to each other before the construction of the \lya\ and continuum images from the NB and BB images (Sec.~\ref{sec:imaging}) for that field. Furthermore, this process ensures that the images across all nine survey fields are also matched to one another before \lya\ or continuum images from different fields are averaged together. The smoothed NB and BB PSFs are very consistent, so we average  the full set of 18 PSF images to construct the final empirical PSF that is assumed for our forward-modeling fits described in Sec.~\ref{sec:mcmcfits}. This final PSF is displayed in Fig.~\ref{fig:finalPSF}.

\end{document}